\documentclass[aps, twocolumn, groupedaddress, nofootinbib, preprintnumbers,superscriptaddress]{revtex4}
\usepackage{float}
\usepackage{aas_macros}
\usepackage[dvips]{graphicx}
\usepackage{color}
\usepackage{amsmath,amssymb,slashed}
\usepackage{tikz}
\usepackage{hyperref}
\usepackage{soul}
\usepackage{tabularx}
\usepackage[normalem]{ulem}

\newcommand{\be}{\begin{equation}}
\newcommand{\ee}{\end{equation}}

\newcommand{\ba}{\begin{eqnarray}}
\newcommand{\ea}{\end{eqnarray}}

\newcommand{\fNL}{f_{\mathrm{NL}}}
\newcommand{\gNL}{g_{\mathrm{NL}}}

\graphicspath{{images/}}

\usepackage{xcolor}

\begin{document}
\title{Where does non-Universality in Assembly Bias come from?}
\author{Charuhas Shiveshwarkar}
\affiliation{Center for Cosmology and AstroParticle Physics (CCAPP), The Ohio State University, 191 West Woodruff Ave, Columbus, OH 43210, USA}
\affiliation{C. N. Yang Institute for Theoretical Physics and Department of Physics \& Astronomy,
Stony Brook University, Stony Brook, NY 11794, USA}
\affiliation{Department of Physics, University of Washington, Seattle, WA 98195, U.S.A.}
\author{Marilena Loverde}
\affiliation{Department of Physics, University of Washington, Seattle, WA 98195, U.S.A.}
\affiliation{C. N. Yang Institute for Theoretical Physics and Department of Physics \& Astronomy,
Stony Brook University, Stony Brook, NY 11794, USA}
\author{Christopher M. Hirata}
\affiliation{Center for Cosmology and AstroParticle Physics (CCAPP), The Ohio State University, 191 West Woodruff Ave, Columbus, OH 43210, USA}
\affiliation{Department of Physics, The Ohio State University, 191 West Woodruff Ave, Columbus, OH 43210, USA}
\affiliation{Department of Astronomy, The Ohio State University, 140 West 18th Avenue, Columbus, OH 43210, USA}
\author{Drew Jamieson}
\affiliation{Max-Planck-Institut für Astrophysik, Karl-Schwarzschild-Straße 1, 85748 Garching, Germany}
\begin{abstract}

Constraints on local primordial non-Gaussianity (LPnG) obtained from galaxy power spectra are limited by the perfect degeneracy between the LPnG parameter $\fNL$ and the bias parameter $b_{\phi}$ which encodes the response of galaxy clustering to a change in the amplitude of primordial curvature fluctuations. For galaxies observed by galaxy surveys, the relation between $b_{\phi}$ and the galaxy bias $b_{g}$ is poorly understood and differs significantly from the universal mass function ansatz. In this paper, we investigate this non-universality in the context of dark-matter halos using the Separate Universe framework, focussing on dark-matter halos selected by mass and/or concentration. We show that the Separate Universe framework provides a natural explanation of the observed universality in the bias of dark-matter halos selected purely by their mass, independent of the spherical collapse picture of halo formation. We further propose an explanation for the observed non-universality in halos selected by concentration and corroborate it with $N$-body simulations in scale-free (EdS) and $\Lambda\text{CDM}$ cosmologies. In particular, we show that the relation between $b_{\phi}$ and halo bias $b_{h}$ for halos selected by concentration in matter-dominated cosmologies tends towards universality at the highest halo masses due to such halos gravitationally dominating their environment throughout their evolution. We also argue that concentration-selected halos of lower masses exhibit non-universality due to their mass accretion being significantly affected by the gravitational influence of neighbouring, more massive halos. Our results suggest that any non-universality in high redshift ($z\gtrsim 3$), high-bias objects observed by realistic galaxy surveys is entirely an artifact of the associated selection function. 
\end{abstract}

\maketitle

\section{Introduction}

Primordial non-Gaussianity (PnG), namely, the existence of higher-point correlations of primordial curvature fluctations (beyond the two-point function), is an important probe of non-standard models of inflation, and provides a window into the physics of the inflationary era~\cite{Meerburg:2019qqi,Achucarro:2022qrl}. The type as well as the extent of PnG are both distinguishing features of different inflationary models~\cite{Meerburg:2019qqi,Achucarro:2022qrl,Babich:2004gb}. Robust measurements of PnG through its impact on Cosmic Microwave Background (CMB) anisotropies and Large-Scale Structure (LSS) can therefore significantly constrain the space of possible inflationary models, thus improving our understanding of the inflationary universe. 

Of particular interest from an observational perspective is PnG of the \textit{local} type, called \textit{local} Primordial non-Gaussianity (LPnG). LPnG is characterised by the existence of non-trivial squeezed limits of one or more higher-point correlation functions~\cite{Bartolo:2004if, Achucarro:2022qrl}, that is, for any $N\geq 3$, we have
\begin{eqnarray}
    \lim_{q\to 0} \frac{\langle\zeta(\textbf{q})\zeta(\textbf{k}_1)\zeta(\textbf{k}_2)...\zeta(\textbf{k}_N)\rangle_{c}}{P_{\zeta}(q)} \neq 0\ ,
\end{eqnarray}
where $\langle\dots\rangle_{c}$ represents a connected correlation function and $P_{\zeta}(q)$ denotes the primordial power spectrum. The existence of a non-trivial LPnG necessitates the existence of at least one light (i.e. $m\ll H$, where $H$ is the inflationary Hubble rate) field in the inflationary universe in addition to the inflaton~\cite{Achucarro:2022qrl,Meerburg:2019qqi,Tanaka:2011aj}. LPnG is thus a smoking gun for multifield inflation and, if detected, would rule out all single-field inflationary models~\cite{Pajer:2013ana,dePutter:2015vga}.

Typical models of LPnG entail primordial curvature fluctuations sourced primarily from the fluctuations of a second light field (called the \textit{curvaton}), with the inflationary expansion driven exclusively by a slowly rolling inflaton~\cite{Enqvist:2001zp, Lyth:2001nq, Sasaki:2006kq,Dvali:2003ar,Dvali:2003em}. In such models, the primordial curvature perturbation $\zeta$ can be written (in real space) as a local polynomial in a Gaussian random field 
\begin{eqnarray}
    \label{Typical_LPnG}
    \zeta(\textbf{x}) &=& \zeta_{G}(\textbf{x}) + \frac{3}{5}f_{\rm NL}\big(\zeta_{G}^{2}(\textbf{x})-\langle\zeta_{G}^2\rangle\big)\nonumber\\
    &&+ \frac{9}{25}g_{\rm NL}\big(\zeta_{G}^{3}(\textbf{x})-3\langle\zeta_{G}^{2}\rangle\zeta_{G}(\textbf{x})\big) +\dots\ ,
\end{eqnarray}
where $\zeta_{G}$ is a Gaussian random field, and the coefficients $\fNL$ and $\gNL$ parametrise the squeezed limits of the primordial bispectrum and the primordial trispectrum respectively. Typical multi-field inflationary models entail $\fNL\sim 1$. A robust detection of $\fNL\sim 1$ is therefore an important observational target for current and upcoming cosmological surveys~\cite{Alvarez:2014vva,Achucarro:2022qrl}. 

The tightest constraints on $\fNL$ and $\gNL$ available today have been obtained from CMB datasets with 68\% confidence limits $\fNL = -0.9\pm 5.1$ and $\gNL = (-5.8\pm 6.5)\times 10^{4}$ obtained by Planck~\cite{Planck:2019kim}. Although the two-dimensional CMB datasets have proved particularly promising in constraining LPnG, any further improvements in their constraining power are likely to be limited by cosmic variance due to the limited number of modes accessible in two-dimensional datasets. LSS surveys provide a three-dimensional dataset with a consequently larger number of accessible modes and provide a promising avenue of improving constraints on LPnG~\cite{Achucarro:2022qrl}. Although constraints on LPnG obtained from LSS datasets so far are not competitive with those obtained by Planck\footnote{see, for example, the constraints on $\fNL$ obtained twelfth data release of the BOSS galaxy survey~\cite{Cabass:2022ymb} or the more recent constraints on $\fNL$ from the scale-dependent bias of the DESI ELG and QSO samples~\cite{Chaussidon:2024qni}.}, current and upcoming LSS surveys such as DESI~\cite{DESI_1}, SPHEREx~\cite{Dore:2014cca}, SpecS5~\cite{SpecS5:2025uom}, and MegaMapper~\cite{Ferraro:2019uce} are expected to have constraining power exceeding that of Planck by almost an order of magnitude~\cite{Dore:2014cca,Ferraro:2019uce}. Indeed, the tightest constraints on LPnG in the future are likely to come from a combination of CMB and LSS datasets~\cite{Schmittful_CMB_lensing,LPnG_kSZ,Quaia}. 
 
LPnG is particularly amenable to detection by LSS surveys as it leads to a large observable signal in the large-scale clustering of galaxies observed at late times~\cite{Achucarro:2022qrl}. A non-trivial squeezed limit of the primordial bispectrum\footnote{Unless otherwise specified, we work with LPnG characterised mainly by a non-trivial primordial bispectrum. The existence of non-trivial higher point primordial correlation functions beyond the bispectrum can also lead to scale-dependent bias effects~\cite{Shiveshwarkar:2023afl,Smith:2010gx,Baumann:2012bc}.} leads to the small-scale matter power spectrum acquiring a non-trivial response to an ambient, large-scale curvature fluctuation, namely,
\begin{eqnarray}
    P_{mm}(\textbf{k}) \rightarrow P_{mm}(\textbf{k})\left(1+\frac{12}{5}\fNL\zeta_{L}\right)\ ,
\end{eqnarray}
where $\fNL$ is a measure of the primordial bispectrum (defined in Eq.~\eqref{Typical_LPnG}.
This modulation of the matter power spectrum also modulates the galaxy abundance by causing the galaxy bias to acquire a scale-dependent correction which increases quadratically (with the inverse wavenumber) out to the largest scales\cite{Neal_fNL_bias}., i.e.
\begin{eqnarray}
    b_{\rm net} &=& b_{g} + \Delta b_{NG}\ ,\\
    \Delta b_{NG} &\propto&\frac{b_{\phi}\fNL}{k^{2}}\ , \label{scale_dependent_bias_def}\\
    b_{\phi} &=& 2\frac{\partial \log n_{g}}{\partial \log A_{s}}\ ,\label{bphi_def}
\end{eqnarray}
where $b_{g}$ is the `Gaussian', scale-independent component of the galaxy bias and $b_{\rm net}$ denotes the total galaxy bias in the presence of LPnG (including the scale-dependent correction). The LPnG bias parameter $b_{\phi}$ encodes the response of tracer abundance $n_{g}$ to a change in the amplitude of primordial curvature fluctuations (parametrised by $A_{s}$). The scale-dependent bias as described in \eqref{scale_dependent_bias_def} is an important observational signal which can be used to constrain the LPnG parameter $\fNL$ using the galaxy power spectrum\cite{Chaussidon:2024qni,Cabass:2022ymb} -- especially because it peaks at the largest scales where the effects of non-linear gravitational clustering are sub-dominant. 

Constraints on $\fNL$ obtained from the scale-dependent bias are limited by the perfect degeneracy between $\fNL$ and $b_{\phi}$~\cite{Barreira:2022sey,Cabass:2022ymb}.\footnote{Using the galaxy bispectrum in addition to the galaxy power spectrum can alleviate this problem to an extent, but at the cost of introducing an additional nuisance parameter $b_{\phi\delta}$ which also occurs in a product combination with $\fNL$~\cite{Cabass:2022ymb} and needs to be modelled. Moreover, $b_{\phi}$ is also a nuisance parameter in modelling galaxy bispectra~\cite{Cabass:2022ymb}.} The non-Gaussian bias parameter $b_{\phi}$ is sensitive to the non-linear physics of galaxy formation and is in general difficult to model from first principles. One therefore needs physically motivated priors on $b_{\phi}$ or some model for the relation of $b_{\phi}$ with the (Gaussian) galaxy bias $b_{g}$ for the scale-dependent bias to be a useful constraining signal for $\fNL$. In this paper we focus on the latter approach of modelling the $b_{\phi}-b_{g}$ relation as a way to improve constraints on $f_{\rm NL}$ that could be obtained from galaxy power spectrum observations. The complementary approach of obtaining physically motivated priors on $b_{\phi}$ (independently of $b_{g}$) proceeds through studies of hydrodynamical simulations of galaxy formation with non-Gaussian initial conditions (see, for example,~\cite{Fondi:2023egm}). Alternatively,~\cite{Dalal:2025eve} and~\cite{Sullivan:2025fie} show how the redshift evolution of the observed number density of tracers could also be used to estimate $b_{\phi}$ for different observed galaxy populations. 

On a related note, it is worth pointing out that one could in principle attempt to constrain the combination $b_{\phi}\fNL$ from galaxy power spectrum measurements. Indeed, even finding $b_{\phi}\fNL\neq 0$ to high statistical significance would still be sufficient evidence to rule out single-field inflationary models. However, the parameter $b_{\phi}$ is generically also redshift dependent making it necessary to model the redshift dependence of $b_{\phi}$ even to rule out $b_{\phi}\fNL=0$\footnote{Galaxy power spectra measured at a single redshift are not sufficiently constraining as to rule out $b_{\phi}\fNL = 0$~\cite{Cabass:2022ymb}.}. On the other hand, a model relating $b_{\phi}$ to the galaxy bias $b_{g}$ is necessary to improve constraints on $\fNL$, especially by leveraging the combined constraining power of LSS and CMB datasets. Additionally, current and upcoming galaxy surveys seek to optimise constraints on $\fNL$ by measuring the clustering of multiple tracers of the matter-density field~\cite{Dore:2014cca,SpecS5:2025uom,DESI_1, Seljak:2008xr} -- each tracer with its own distinctive $b_{\phi}-b_{g}$ relation. Constraints on $\fNL$ obtained in this way can also be particularly sensitive to different models of the $b_{\phi}-b_{g}$ relationship for different tracers~\cite{Barreira_Krause}. 

Tracers whose abundance can be expressed solely as a function of the amplitude of matter clustering $\sigma_{8}$, and not dependent on any other cosmological parameters are referred to as \textit{universal}. For such tracers, it has been found that $b_{\phi}$ bears a simple relation with the scale-independent galaxy bias\cite{Neal_fNL_bias}:
\begin{eqnarray}
    b_{\phi} = (b_{g}-1)\delta_{c}\ ,\label{universality_def}
\end{eqnarray}
where $b_{g}$ is the (Eulerian) scale-independent galaxy bias and $\delta_{c}$ is the threshold overdensity for spherical collapse, namely $\delta_{c}=1.686$. Throughout this paper, we refer to the violation of the relation~\eqref{universality_def} as \textit{non-universality} and tracers which violate Eq.~\eqref{universality_def} are consequently \textit{non-universal}. The universality ansatz \eqref{universality_def} has been verified to be true to a significant accuracy for dark matter halos selected purely by mass.\footnote{The existence of higher-order LPnG, i.e. the existence of non-trivial squeezed limits of higher-point correlation functions of primordial curvature fluctuations will necessary violate universality~\cite{LPnG_HMF,Shiveshwarkar:2023afl}. Even in this case, for dark-matter halos selected by mass alone, universality is only weakly broken~\cite{LPnG_HMF}.} However, tracers of matter density observed by galaxy surveys as well as dark matter halos selected by properties beyond mass typically show significant deviations from universality\cite{Desjacques:2016bnm, Barreira:2022sey, Barreira:2020kvh, Lazeyras:2022koc}. Most such deviations from the universality relation \eqref{universality_def} can be encoded in a single parameter $p_{NG}$ as  
\begin{eqnarray}
    b_{\phi} = (b_{g}-p_{NG})\delta_{c}\ ,\label{p-def}
\end{eqnarray}
where $p_{NG}$ is an $\mathcal{O}(1)$ parameter but is in general different from 1.0. For example, \citet{Barreira:2020kvh} find that galaxies selected by stellar mass typically show $p_{NG} \in [0.5,0.7]$ and \citet{Slosar_2008} show that quasars typically exhibit $p_{NG}\sim 1.6$. For dark matter halos, \citet{Lazeyras:2022koc} show that the value of $p_{NG}$ is significantly different from 1.0 for halos selected by mass and concentration with higher concentration halos typically showing a lower value of $p_{NG}$ (and consequently higher $b_{\phi}$) than halos with lower concentrations. On the other hand, \citet{Lazeyras:2022koc} also show that $p_{NG}$ for halos selected by mass and sphericity or by mass and spin is $\sim 1$ -- i.e. there is no significant non-universality in the assembly bias of halos selected by mass and sphericity or by mass and spin. 

The case of halos selected by mass and concentration is thus interesting as a test case which can shed light on the deeper origin of deviations from non-universality. In this paper, we use the Separate Universe framework to investigate the physical reason behind the non-universal clustering of dark matter halos selected by mass and concentration.
\newline
\indent The rest of the paper is organised as follows. In section \ref{sec:Halo bias and the Separate Universe}, we present our approach for deriving the relation between $b_{\phi}$ and the halos bias $b_{h}$ using the Separate Universe framework. In particular, we show that our approach can explain the universality in the clustering of halos selected purely by mass. Along the same lines, we also present our argument for why universality should be restored for halos selected by mass and concentration at high halo masses. Section \ref{sec:Methods} reviews details about the N-body simulations used in this paper as well as our techniques for measurement of $b_{h}$ and $b_{\phi}$ for halos selected by mass and concentration. We present the results of our N-body simulations in section \ref{sec:Results} and show that they are consistent with our hypotheses. We summarise our results in section \ref{sec:Discussion} and also discuss the effect of tracer selection function on the $b_{\phi}-b_{g}$ relation.

\section{Halo bias and the Separate Universe}
\label{sec:Halo bias and the Separate Universe}
\indent The Separate Universe framework entails that the effect of a long-wavelength matter (i.e. Cold Dark Matter (CDM) + Baryon\footnote{Unless otherwise specified, by `matter density fluctuations' we mean the density constrast of CDM + Baryons.}) fluctuation can be understood in terms of a reparametrisation of the background cosmology to yield a set of \textit{local} cosmological parameters which govern the growth of structure within the large-scale fluctuation~\cite{Hu:2016ssz}. The large-scale bias of galaxies is then identical to the response of the galaxy number density to a long-wavelength matter density fluctuation computed in this framework~\cite{SU_consistency}. In other words, 
\begin{eqnarray}
\lim_{k\rightarrow 0} \frac{P_{gm}(k,z)}{P_{mm}(k,z)} = 1+\frac{d\log n_{g}}{d\delta_{L}}\ ,\label{SU_bias_identity}
\end{eqnarray}
where $P_{mm}$ and $P_{gm}$ are the matter power spectrum and the galaxy-matter cross power spectrum respectively whereas the right hand side is the response of the mean (comoving)\footnote{Unless otherwise specified, we work with comoving number densities in this paper.} number density of galaxies $n_{g}$ to an ambient large-scale matter density fluctuation $\delta_{L}$. 

The formation of galaxies or dark matter halos is a `quasi-local' process -- in the sense that the distance-scales relevant for halo/galaxy formation are much smaller than typical scales at which one studies their clustering~\cite{Desjacques:2016bnm} -- and is governed by the \textit{small-scale} properties of the matter fluctuation fields -- namely the matter density field, tidal field, peculiar velocity field, linear growth rate, etc. The mean number density of dark matter halos and galaxies should thus be a function of one or more small-scale statistics of said matter fluctuation fields, i.e. $n_{g} = n_{g}(\{\theta_{i}\})$ where each of the $\theta_{i}$ is a statistic constructed from the small-scale matter fluctuation fields and/or their derivatives. The simplest such statistics are the amplitude of matter fluctuations $\sigma_{M}(z)$ smoothed over a mass scale $M$ corresponding to the typical masses of dark matter halos, or the linear growth rate $\alpha_{\text{eff}} = d\log\delta_{m}/d\log a$, or an effective tilt in the small-scale matter power spectrum parameterised by $n_{\text{eff}} = d\log \sigma_{M}/d\log M$.

 The separation of scales in the Separate Universe construction~\cite{Hu:2016ssz} (together with the quasi-local nature of halo formation) implies that an ambient large-scale matter density affects the mean abundance of dark matter tracers $n_{g}$ \textit{only} through a modulation of the small-scale statistics $\theta_{i}$. In other words, 
\begin{eqnarray}
    b_{L} = \frac{d\log n_{g}}{d\delta_{L}} = \sum_{i}\frac{\partial \log n_{g}}{\partial \theta_{i}}\frac{d\theta_{i}}{d\delta_{L}}\ ,\label{SU_bias_expression}
\end{eqnarray}
where $b_{L}$ is the lagrangian bias of galaxies and the responses $d\theta_{i}/d\delta_{L}$ of the small-scale statistics $\theta_{i}$ can be naturally computed within the Separate Universe framework. One the other hand, $b_{\phi}$ can be expressed in a similar way as,
\begin{eqnarray}
    b_{\phi} = 2\frac{d\log n_{g}}{d\log A_{s}} = \sum_{i}\frac{\partial\log n_{g}}{\partial \theta_{i}}\cdot2\frac{d\theta_{i}}{d\log A_{s}}\ ,\label{SU_bphi_expression}
\end{eqnarray}
where $A_{s}$ is the amplitude of primordial curvature power spectrum ($P_{\zeta} \propto A_{s}$). Equations \eqref{SU_bias_expression} and \eqref{SU_bphi_expression} show that the relation between $b_{g}$ and $b_{\phi}$ is governed by the \textit{difference} between how the relevant small-scale statistics $\theta_{i}$ of the matter fluctuation fields respond to an ambient large-scale \textit{matter density} field and an increase in the amplitude of primordial curvature fluctuations. The problem of modelling the connection between $b_{\phi}$ and $b_{g}$ for a class of matter density tracers therefore maps on to the problem of determining which small-scale statistical measures \{$\theta_{i}$\} of the matter density field are needed to completely characterise the abundance of said matter density tracers and modelling their response to a large-scale matter density fluctuation and to an increase in the amplitude of primordial curvature fluctuations respectively\footnote{Note that at leading order, an increase in $A_{s}$ merely changes the initial conditions for structure formation whereas a large-scale matter density fluctuation $\delta_{L}$ primarily changes the time evolution of cosmological density perturbations (when appropriately defined w.r.t the reference frame of a local observer within the large-scale density fluctuation~\cite{Hu:2016ssz,Chiang:2018laa,Shiveshwarkar:2020jxr}). } . Universality in this language is the statement that all the relevant statistics $\{\theta_{i}\}$ in Eq.s~\eqref{SU_bias_expression} and~\eqref{SU_bphi_expression} obey :
\begin{eqnarray}
    \frac{d\theta_{i}}{d\delta_{L}}\cdot\delta_{c} = 2\cdot\frac{d\theta_{i}}{d\log A_{s}}\ .\label{universality_statistics}
\end{eqnarray}

On the other hand, non-universal halo clustering necessarily implies that halo abundance has a significant dependence on late-time properties of the matter density field which \textit{violate} Eq.~\eqref{universality_statistics}. Examples of such quantities include the halo-merger rate at late times~\cite{Slosar:2008hx}, or the tidal anisotropy~\cite{Ramakrishnan:2019wtt} of the halo environment, or the scale-independent growth rate $f(z) = d\log D/d\log a$ (for non-Einstein-de Sitter cosmologies).
\newline
\subsection{Halos selected by mass}
\label{ssec:Halos selected by mass}
\indent In Einstein-de Sitter (EdS) and typical $\Lambda\text{CDM}$ cosmologies (without LPnG), the abundance of halos of a certain mass $M$ is a function only of the mass variance $\sigma(M,z)$ (or alternatively, the peak-height $\nu(M,z)= \delta_{c}/\sigma(M,z)$) smoothed on the Lagrangian scale of the mass $M$~\cite{PressSchechter1974,Bardeen:1985tr, Kaiser:1987qv,ePS1, ePS2, Cooray:2002dia}. This follows from the statistics of peaks in the primordial matter density field which collapse to form dark matter halos at late times~\cite{Bardeen:1985tr, Kaiser:1987qv,ePS1, ePS2}. This is a special case of equations~\eqref{SU_bias_expression},~\eqref{SU_bphi_expression} where the only relevant statistic needed to characterise halo abundance completely is the mass variance $\sigma(M,z)$ smoothed on a certain mass scale at a certain redshift. Following equations~\eqref{SU_bphi_expression} and~\eqref{SU_bias_expression}, we have (for halos selected purely by mass) :
\begin{eqnarray}
    b_{L}(M,z) &=& \frac{d\log n(M,z)}{d\delta_{L}}\ ,\nonumber \\
               &=& \frac{d\log n}{d\log \sigma_{M}}\cdot\frac{d\log\sigma_{M}}{d\delta_{L}}\ ,\nonumber \\
               &=& 2\cdot\frac{d\log n}{d\log A_{s}}\cdot R_{SU}(M,z)\ ,\label{SU_mass_selected_halos}
\end{eqnarray}
where $R_{SU}$ is the response of $\sigma_{M}$ to an ambient large-scale scale matter density $\delta_{L}$ computed within the Separate Universe framework. For EdS cosmologies, $R_{SU} = 13/21$ at leading order and is independent of mass and redshift~\cite{Hu:2016ssz,Chiang:2017vuk, Chiang:2014oga}. For $\Lambda\text{CDM}$ cosmologies, $R_{SU}\approx 13/21$ at leading order until the end of the matter dominated epoch and is only weakly redshift-dependent thereafter while still remaining independent of mass~\cite{Hu:2016ssz}. In either case, for both EdS and $\Lambda\text{CDM}$ cosmologies, Eq.~\eqref{SU_mass_selected_halos} yields the following relation between the (Eulerian) halo bias $b_{h}(M,z)$ and $b_{\phi}(M,z)$ for halos selected purely by mass :
\begin{eqnarray}
    b_{\phi} = (b_{h}-1)R^{-1}_{SU}\ .\label{universality_mass_selected_halos}
\end{eqnarray}
In both EdS and typical $\Lambda\text{CDM}$ cosmologies, $R^{-1}_{SU} \sim 1.6 \sim \delta_{c}$ and is independent of mass and only weakly redshift-dependent. Note that Eq.~\eqref{universality_mass_selected_halos} reproduces the standard form of the universality relation (Eq.~\eqref{universality_def}) to within $\sim 6\%$.
\newline
\indent N-body simulations with different halo-finding strategies and different halo mass definitions often find that a slightly modified version of the universality relation Eq.~\eqref{universality_def}/Eq.~\eqref{universality_mass_selected_halos} holds for halos selected by mass, namely $b_{\phi} = (b_{h}-1)q\delta_{c}$ where $q\delta_{c}\approx 1.42$~\cite{Biagetti:2016ywx, Hadzhiyska:2025rez}. Similarly, we find that halos identified by the Friends-of-Friends (FOF) algorithm but with masses assigned using a spherical overdensity (SO) mass-assignment scheme~\cite{RockstarPaper} and selected by their virial masses in our N-body simulation obey a universality relation of the form given by $\delta_{c} \approx 1.42$ instead of $\delta_{c}\sim 1.6$ (which would be expected from $R^{-1}_{SU}\sim 1.6$). This is consistent with the findings of~\cite{Biagetti:2016ywx} where they find that FOF halos (with SO masses) selected by mass obey a universality relation with $b_{\phi}/\delta_{c}b_{L}\sim 0.89$, which is equivalent to $\delta_{c} \approx 1.42$.

\subsection{Halos selected by mass and concentration}
\label{ssec:Halos selected by mass and concentration}
\indent Halo concentration is tightly correlated with halo formation time which in turn depends on the mass accretion history of the halo~\cite{Wechsler2002,Correa2015}. Halos with higher mass accretion rates have a later `formation' epoch (\cite{Correa2015}) beyond which halo concentration primarily grows through an increase in the virial radius ($R_{\text{vir}}$) while the scale radius ($r_{s}$) of the halo remains nearly constant~\cite{Wechsler2002,Correa2015, Diemer:2018vmz}. As a result, halos with a higher mass accretion rate typically have \textit{lower} observed concentration than their counterparts with lower mass accretion rates~\cite{Correa2015, Wechsler2002}. 

The assembly bias of halos selected by mass and concentration shows opposite trends at high and low halo masses (compared to the non-linear mass scale $M_{*}$ defined by $\nu(M,z)=1$). At the high-mass ($M\gg M_{*}$) end, halo bias tends to \textit{decrease} with increasing halo concentration. High-mass $(M\gg M_{*})$ halos are formed from the collapse of large peaks in the primordial matter density field~\cite{Bardeen:1985tr, PressSchechter1974, ePS1, ePS2}. The decreasing trend of the bias of such halos w.r.t concentration is a consequence of the statistics of large peaks in the (nearly) Gaussian primordial matter density field~\cite{Bardeen:1985tr, Dalal:2008zd,Kaiser:1987qv}.  At the low-mass ($M\ll M_{*}$) end however, this trend is reversed -- halos with a higher concentration have \textit{higher} biases than their low-concentration counterparts~\cite{TidalEffects,Ramakrishnan:2019wtt}. 

Several explanations have been proposed (see, for example~\cite{TidalEffects,TidalEffects2,Dalal:2008zd}) for this inversion in the trend of halo assembly bias w.r.t concentration at low halo masses. These explanations typically rely on the fact that the mass accretion rates of low-mass ($M\ll M_{*}$) halos (unlike those of their high-mass counterparts) are significantly affected by properties of their immediate small-scale environment\footnote{By the halo `environment' we mean the small-scale configuration of a proto-halo patch within which the halo forms. The size of this patch is typically much smaller than the scales at which one might compute clustering properties such as the bias; in line with the quasi-local nature of halo formation.} -- especially the tidal effects of neighbouring, more massive halos~\cite{TidalEffects,TidalEffects2,Dalal:2008zd,Ramakrishnan:2019wtt}. Tidal stripping due to the influence of a neighbouring high-mass ($M\gg M_{*}$) halo causes the mass accretion of a low-mass halo to be prematurely truncated leading to the low-mass halo having a \textit{higher} concentration. Moreover, low-mass halos formed in this way (i.e. by escaping accretion into high-mass halos) also have a \textit{higher} bias than their more isolated counterparts because they are preferentially formed in high matter-density regions -- as evidenced by the presence of neighbouring high-mass peaks. The gravitational influence of high-mass halos thus creates a sub-population of \textit{high-concentration} and \textit{high-bias} low-mass halos~\cite{Dalal:2008zd,Ramakrishnan:2019wtt} which leads to the well-known inversion in the trend of assembly bias w.r.t concentration at low halo mass. 

In this subsection, we propose an explanation of the non-universal relation between halo bias $b_{h}$ and $b_{\phi}$ observed in concentration-selected halos along similar lines. We consider the cases of high-mass (with $M \gg M_{*}$) and low-mass (with $M\ll M_{*}$) halos separately.

\subsubsection{High mass ($M\gg M_{*}$) halos selected by concentration}
\label{sssec:High mass halos selected by concentration}

\indent As mentioned earlier, halos with high mass ($M\gg M_{*}$) form as a result of the nearly spherical collapse of large, rare peaks in the primordial matter density field~\cite{Bardeen:1985tr, PressSchechter1974, ePS1, ePS2}. Large peaks in the primordial matter density field dominate their immediate environment -- with the result that the mass accretion rate of the highest mass halos is nearly independent of the small-scale environment in which the halo is formed and is only dependent on the properties of the initial peak. As a result, the concentration of a high-mass halo is also strongly correlated to properties of the initial peak. In particular, the curvature of the initial density peak, $\xi=R^{2}\nabla^{2}\delta_{R}$ (smoothed on the lagrangian scale $R$ of a high-mass halo) has been shown to be significantly correlated to the observed concentration of the high-mass halo~\cite{Dalal:2008zd}\footnote{Note that~\cite{Dalal:2008zd} use a slightly different, but equivalent parametrisation of the peak curvature.}. Halos formed from highly curved peaks\footnote{Note that highly curved peaks correspond to peak with a more negative (and thus, \textit{lower}) peak curvature $R^{2}\nabla^{2}\delta_{R}$ !} have an earlier `formation' time (consistent with a lower mass accretion rate) and consequently, a higher concentration~\cite{Dalal:2008zd} than halos formed through the collapse of flatter peaks.\footnote{In principle, the concentration of high-mass peaks can be affected by tidal stripping due to the gravitational influence of massive neighbours, but high-mass ($M\gg M_{*}$) peaks are also `rare' peaks (i.e. much less likely to be in proximity to a more/comparably massive peak)~\cite{Dalal:2008zd} and dominate their immediate environment -- so that tidal stripping from the influence of neighbouring peaks gives a subdominant contribution to their concentration. } This connection between concentration and peak curvature can adequately explain the assembly bias of high-mass halos selected by mass and concentration in terms of the statistics of peaks in the (nearly gaussian) primordial matter density field (selected by their peak-height and peak curvature)~\cite{Dalal:2008zd, Bardeen:1985tr}.  

\indent In terms of the Separate Universe picture of halo bias and $b_{\phi}$ (outlined in section~\ref{sec:Halo bias and the Separate Universe}), the connection between peak curvature and halo concentration implies that the abundance of halos selected by both mass and concentration in pure EdS cosmologies is a function not just of the mass variance $\sigma_{R}$ but also the variance of peak curvature\footnote{Note that $R$ here is the lagrangian radius corresponding to halos of a given mass.} $\sigma^{2}_{\xi,R} = \langle \left(R^{2}\nabla^{2}\delta_{R}\right)^{2}\rangle$ and the cross-correlation coefficient $\gamma_{R} = \langle\delta_{R}\cdot R^{2}\nabla^{2}\delta_{R}\rangle/\sigma_{R}\sigma_{\xi,R}$. Following equations~\eqref{SU_bphi_expression} and~\eqref{SU_bias_expression} we have the following expressions for the (lagrangian) halo bias $b_{L}$ and $b_{\phi}$ of halos of mass $M$ and concentration $c$ (corresponding to a peak curvature $\xi$) :
\begin{eqnarray}
b_{L}(M,c) &=& \frac{\partial \log n}{d\delta_{L}}\ ,\nonumber\\
&=& \frac{\partial \log n}{\partial\log \sigma_{R}}\frac{\partial\log\sigma_{R}}{\partial\delta_{L}}\nonumber\\
&&+ \frac{\partial \log n}{\partial\log \sigma_{\xi,R}}\frac{\partial\log\sigma_{\xi,R}}{\partial\delta_{L}}\nonumber\\
&&+ \frac{\partial \log n}{\partial \gamma_{R}}\frac{\partial\gamma_{R}}{\partial\delta_{L}}\ ,\label{SU_bias_high_Mass_concentration}
\end{eqnarray}
and 
\begin{eqnarray}
b_{\phi}(M,c) &=& 2\frac{\partial \log n}{d\log A_{s}}\ ,\nonumber\\
&=& \frac{\partial \log n}{\partial\log \sigma_{R}}\cdot 2\frac{\partial\log\sigma_{R}}{\partial\log A_{s}}\nonumber\\
&&+ \frac{\partial \log n}{\partial\log \sigma_{\xi,R}}\cdot 2\frac{\partial\log\sigma_{\xi,R}}{\partial\log A_{s}}\nonumber\\
&&+ \frac{\partial \log n}{\partial \gamma_{R}}\cdot2\frac{\partial\gamma_{R}}{\partial\log A_{s}}\ .\label{SU_bphi_high_Mass_concentration}
\end{eqnarray}

It can be shown (see Appendix~\ref{app:responses}) that to leading order,
\begin{equation}
    \label{gamma_r_responses}
    \frac{\partial \gamma_{R}}{\partial\delta_{L}} = \frac{\partial\gamma_{R}}{\partial\log A_{s}} = 0\ ,
\end{equation}
and 
\begin{eqnarray}
    \label{peak_curvature_responses}
    \frac{\partial\log\sigma_{\xi,R}}{\partial\delta_{L}} = 2\frac{\partial\log\sigma_{\xi,R}}{\partial\log A_{s}}\cdot R_{SU}(M,z)\ .
\end{eqnarray}
Following the results of subsection~\ref{ssec:Halos selected by mass}, it follows that for halos selected by mass and concentration in the high-mass regime (where selecting halos by concentration is equivalent to selecting halos by their initial peak curvature), $b_{L}$ and $b_{\phi}$ obey the universality relation with $p_{NG}=1$; i.e 
\begin{equation*}
    b_{L}(M,c) = b_{\phi}(M,c)R_{SU} \approx b_{\phi}(M,c)\delta^{-1}_{c} .
\end{equation*}
In other words, the $b_{L}-b_{\phi}$ relation for halos selected by mass and concentration in a purely matter dominated cosmology should approach universality in the high-mass regime. 

As a corollary of the argument presented above, it follows that halos selected by properties which are tightly correlated to the properties of the corresponding initial peak in the linear density field should obey the universality relation. This is because the primordial density field is very nearly gaussian and the only relevant statistical properties of an individual peak that can be constructed are mass variance $\sigma_{R}$ or the derivatives thereof -- all of which obey~\eqref{universality_statistics}. In particular, halo abundances modelled using traditional Press-Schechter/extended-Press-Schechter approaches are, by construction, exclusively dependent on the mass-variance $\sigma_{M}$ or derivates thereof and are therefore manifestly \textit{universal}.


\indent The arguments in this section really only hold in a purely matter-dominated universe. In typical $\Lambda\text{CDM}$ universes, the median halo concentration also depends on the scale-independent linear growth rate (~\cite{Diemer:2018vmz, DiemerK}) $f(z)$ which tends to be significantly different from $1$ at late times (in fact $f\sim 0.5$ at $z=0$) -- contributing to the non-universality of the $b_{\phi}-b_{L}$ relationship. However, as we show in section~\ref{ssec:Results for Halos selected by mass and concentration} (figure~\ref{fig:LCDM_bphi_vs_b0_all_nu}), the non-universality in the $b_{L}-b_{\phi}$ relation for halos of the same mass does tend to decrease with redshift as would be expected from the fact that the universe becomes more and more matter-dominated and the non-linear mass scale rapidly decreases (so that halos of a given mass eventually approach a regime where $M \gg M_{*}$ and their concentration starts getting more correlated to their respective peak curvatures).

\subsubsection{Low mass ($M\ll M_{*}$) halos selected by concentration}
\label{sssec:Low mass halos selected by concentration}

\indent At significantly low halo masses, halo concentration becomes more correlated with the environment in which the halo was formed (in terms of proximity to large peaks in the matter density field) than the Lagrangian properties of the corresponding initial peak~\cite{Dalal:2008zd,Ramakrishnan:2019wtt}. Low-mass halos that form from the collapse of small peaks that are in proximity to a larger, more massive peak have their mass accretion truncated due to tidal stripping from the gravitational influence of the neighbouring massive peak~\cite{Dalal:2008zd}. Such halos have an earlier `formation' time and present with high observed concentrations~\cite{Dalal:2008zd}. On the other hand, low-mass halos that form from isolated peaks in the primordial density fields do not have their mass accretion truncated prematurely and present with a lower observed concentration~\cite{Dalal:2008zd}. That is why in the low-mass limit (i.e. $M\ll M_{*}$), segregating halos by concentration is a good proxy for their segregation by proximity to more massive peaks (or equivalently, by the tidal anisotropy of their immediate environment~\cite{Ramakrishnan:2019wtt}). This means the abundance of halos segregated in different quantiles of halo concentration is governed not just by the lagrangian properties of their respective progenitor peaks but also by properties of the matter-density field in their immediate environment\footnote{Note, in particular, that the concentration of low-mass, low-concentration halos is still largely dependent on their peak curvature -- unlike their abundance which depends on properties beyond those of their progenitor peaks.} -- which are essentially \textit{non-universal} in the sense of Eq.~\eqref{universality_statistics}.

 \indent ~\citet{Dalal:2008zd} present a toy model incorporating the above connection between concentration and bias of low-mass halos to explain the inversion in the trend of assembly bias w.r.t concentration at low halo masses. In this section we propose a heuristic toy model built along similar lines which, under general assumptions about the nature of halo formation and their connection with statistics of peaks in the primordial density field, can qualitatively explain the deviation from universality exhibited by low-mass halos selected by concentration. The low-mass peaks which collapse to form halos of a given mass $M$($\ll M_{*}$) can be divided into two sub-populations -- a fraction $\eta$ of peaks which are in proximity to large, more massive peaks and have their mass accretion truncated prematurely (called population $\text{I}$ in what follows) and the remaining fraction ($1-\eta$) of peaks which are isolated and don't have their mass accretion prematurely stunted (called population $\text{II}$ in what follows). Peaks of the former kind evolve to form halos if they escape accretion into the neighbouring larger mass halo -- the surviving peaks have their growth stunted and produce a population of older, more concentrated halos~\cite{Dalal:2008zd}. The survival probability $\chi$ of such peaks is correlated with their binding energy relative to the neighbouring, more massive peak~\cite{Dalal:2008zd}. The number density of low-mass halos can therefore be written as 
\begin{eqnarray}
    n_{h}(M,z) &=& n_{\text{I}} + n_{\text{II}}\nonumber\ ,\\
    &=& n_{pk}(\nu)\eta\chi + n_{pk}(\nu)(1-\eta) ,\label{halo_abundance_low_masses}
\end{eqnarray}
where $n_{pk}(\nu)$ is the number density of low-mass peaks which are the ancestor peaks of low-mass halos of mass $M$ and $n_{\text{I}} = n_{pk}\eta\chi$ and $n_{\text{II}} = n_{pk}(1-\eta)$ are the number densities of halos in population $\text{I}$ (i.e. older, higher concentration halos) and population $\text{II}$ (i.e. newer, lower concentration halos) respectively. Following the connection between halo abundance and the abundance of peaks in the primordial density field,\footnote{see, for examples~\cite{Bardeen:1985tr,PressSchechter1974}.} $n_{pk}$ is a function only of the peak-height $\nu$ corresponding to the mass $M$ at redshift $z$ and therefore obeys the `universality' condition (in the sense of equation~\eqref{universality_mass_selected_halos}) :
\begin{eqnarray}
    \frac{\partial \log n_{pk}}{\partial\delta_{L}} = 2\cdot\frac{\partial\log n_{pk}}{\partial\log A_{s}}R_{SU}\ .\label{general_universality}
\end{eqnarray}
On the other hand, the escape probability $\chi$, and the relative probability of being in proximity to more massive peaks $\eta$ would in general be \textit{non-universal} (in the sense that Eq.~\eqref{general_universality} would not hold for $\eta$ and $\chi$). Note that $\chi$ here is the \textit{conditional} escape probability -- i.e. the probability that a low-mass ($M\ll M_{*}$) peak forms a halo \textit{given} that it is in proximity to a large mass ($M\gg M_{*}$) peak. In the limit of a large number of low-mass peaks (which is true for the low-mass limit $M\ll M_{*}$) which are either isolated or in proximity to larger peaks, one can reasonably assume that $\chi$ is also the \textit{conditional} escape fraction -- i.e. the fraction of all those low-mass peaks which are in proximity to larger peaks that form halos observable at late times.

Going forward, we analyse how the conditional escape probability $\chi$ responds to a change in the amplitude of primordial fluctuations $A_{s}$ and to a background long-wavelength matter density fluctuation $\delta_{L}$. Since the presence of a background long-wavelength matter density is, to leading order, equivalent to a positive spatial curvature~\cite{Hu:2016ssz}, one would expect $\chi$ to \textit{decrease} in the presence of an ambient long-wavelength matter fluctuation, leading to\footnote{Another way of seeing this is to note that all particle geodesics get a convergent drift in the presence of spatial curvature~\cite{MTW} making it less likely for a small matter density peak to escape from accretion into a neighbouring massive peak in the presence of a background long-wavelength matter overdensity.} 
\begin{eqnarray}
    \label{chi_response}
    \frac{\partial\chi}{\partial\delta_{L}} < 0\ .
\end{eqnarray}
To intuitively understand the dependence on $A_{s}$, we note that the conditional escape probability $\chi$ is largely sensitive to the dynamics of gravitational collapse of the low-mass peak and its high-mass neighbour which in turn is dependent only on the gravitational interaction between the low-mass peak and its high-mass neighbour. The latter is primarily a function of the masses of the proto-halo patches surrounding both the peaks and the physical distance between them -- quantities which are not significantly affected by changes in $A_{s}$. One would therefore expect that the conditional escape probability $\chi$ is relatively independent of $A_{s}$ and obeys 
\begin{eqnarray}
    \label{A_s_response}
    \frac{\partial\chi}{\partial\log A_{s}} \approx 0\ .
\end{eqnarray}

A more insightful way to see why the conditional escape probability $\chi$ should satisfy Eq.~\eqref{A_s_response} and Eq.~\eqref{chi_response} is as follows. We make the reasonable assumption (following the fact that $\chi$ is largely correlated with the initial binding energy of a low-mass peak~\cite{Dalal:2008zd}) that $\chi$ is a function of the initial gravitational binding energy of the low-mass peak (computed at a sufficiently early time before the epoch of halo-formation). This initial binding energy of the low-mass peak is dominated by the two-body gravitational interaction with its high-mass neighbour. To obtain the Separate Universe response of the binding energy of the low-mass peak to an ambient long-wavelength fluctuation, we compute it in the `physical' Fermi coordinates centred at the trajectory of the neighbouring high-mass peak\footnote{Note that this is justified because (a) the evolution of the high-mass peak very nearly follows the spherical collapse model and is nearly unaffected by the gravitational influence of any low-mass neighbours~\cite{Dalal:2008zd} and (b) The physical distance between the low-mass peak and its high-mass neighbour is still within the Hubble radius at the sufficiently early time ($z\sim 49$; before the epoch of halo-formation when the density fluctuations on relevant scales evolve linearly).}. We use Fermi coordinates because they form a locally inertial, physical coordinate system corresponding to the reference frame of a freely falling small-scale observer (along the entire length of the observer's trajectory) and are a natural choice of coordinates to obtain the physical response of local (sub-horizon) observables to ambient long-wavelength perturbations~\cite{FNC_1,FNC2}. The binding energy of the low-mass peak computed in these coordinates has a leading order component $\mathcal{E}_{\rm FRW}=\frac{3}{4}H^{2}\textbf{x}^{2}$ corresponding to the Hubble flow around the massive peak in the matter-dominated epoch \footnote{Note that $\mathcal{E}_{\rm FRW}$ includes both the potential energy $\Phi_{\rm FRW} = \frac{1}{4}H^{2}\textbf{x}^{2}$ and the kinetic energy $T_{\rm FRW} = \frac{1}{2}\textbf{v}^{2}_{FRW}$ where $\textbf{v}_{\rm FRW} = H\textbf{x}$ associated with the Hubble flow. See appendix~\ref{app:FNC} for more details. }reflecting the global expansion of the unperturbed universe (see Appendix~\ref{app:FNC}) and subleading terms arising from fluctuations around the FRW background which would consequently be suppressed by powers of $A^{1/2}_{s}D_{\rm ini}$ where $D_{\rm ini}$ is the initial linear growth factor (this sub-leading contribution includes both the kinetic energy contribution due to peculiar velocities as well as the gravitational potential in the perturbed FRW cosmology -- both are $\propto A^{1/2}_{s}D_{\rm ini}$ at sufficiently early times). In other words, the binding energy can be written as 
\begin{eqnarray}
    \mathcal{E} = \mathcal{E}_{\rm FRW} + \Delta\mathcal{E}\label{FNC_binding_energy}\ ,
\end{eqnarray}
where $\mathcal{E}_{\rm FRW} = \frac{3}{4}H^{2}\textbf{x}^2$ (with $H$ being  the Hubble rate at the early time) and $\Delta\mathcal{E}\sim\mathcal{O}(A^{1/2}_{s}D_{\rm ini})$.

From Eq.~\eqref{FNC_binding_energy}, it follows that the Separate Universe response of the binding energy is dominated by the response of the Hubble rate~\cite{Hu:2016ssz} $H\rightarrow H(1-\frac{1}{3}\delta_{L})$ (with sub-dominant terms suppressed by powers of $A^{1/2}_{s}D_{\rm ini}$)\footnote{Note that `physical' coordinates in the Separate Universe are the same as those in the Global Universe~\cite{Hu:2016ssz}.}. In comparison, the response $\partial\chi/\partial\log A_{s}\sim \Delta\mathcal{E}/\mathcal{E}_{\rm FRW}$ is $\sim \mathcal{O}(A^{1/2}_{s}D_{\rm ini})$. Since a more negative binding energy implies a reduced conditional escape probability $\chi$~\cite{Dalal:2008zd}, we have, to leading order in $A^{1/2}_{s}$,
\begin{eqnarray}
    \frac{\partial\log\chi}{\partial\delta_{L}} &=& \frac{d\log\chi}{d\mathcal{E}}\cdot\frac{\partial\mathcal{E}}{\partial\delta_{L}} < 0\ ,\\
    \frac{\partial\log\chi}{\partial\log A_{s}} &=& \frac{d\log\chi}{d\mathcal{E}}\cdot\frac{\partial\mathcal{E}}{\partial\log A_{s}} \approx 0\ .
\end{eqnarray}

\indent Having qualitatively analysed the variation of the conditional escape probability w.r.t a background long-wavelength matter-density and w.r.t a change in $A_{s}$, we now make contact with the Separate Universe picture of universality (see section~\ref{sec:Halo bias and the Separate Universe}). The results of our simulations~\ref{ssec:Results for Halos selected by mass} as well as the arguments of subsection~\ref{ssec:Halos selected by mass} imply that the total number density of low-mass halos (i.e. $n_{h}(M,z)$ in Eq.~\eqref{halo_abundance_low_masses} obeys the universality condition (Eq.~\ref{general_universality}). This implies that the non-universalities in the fraction $\eta$ and the escape probability $\chi$ should cancel out, leading to the following consistency condition :
\begin{eqnarray}
    \frac{\partial\log\eta}{\partial\delta_{L}} - 2\frac{\partial\log \eta}{\partial\log A_{s}}R_{SU} = \frac{\chi}{1-\chi}\frac{\partial\log\chi}{\partial\delta_{L}}\ .\label{consistency_constraint}
\end{eqnarray}
Using the above consistency relation we can proceed to compute the non-universality in low-mass, high-concentration halos (i.e. population $\text{I}$) and for low-mass, low-concentration halos (i.e. population $\text{II}$) in terms of $p_{NG}$ (see Eq.~\eqref{p-def}) where we can express $p_{NG}$ as :
\begin{eqnarray}
    p_{NG} = 1 + \frac{\partial\log n}{\partial\delta_{L}} - 2\frac{\partial\log n}{\partial\log A_{s}}\cdot R_{SU}\ ,\label{p-SU-expression}
\end{eqnarray}
where we have used Eq.~\eqref{SU_bias_identity} to express the Eulerian halo bias in terms of the Separate Universe response of halo abundance. Following equation~\eqref{p-SU-expression}, we obtain the following expressions for the values of $p_{NG}$ for the high- and low-concentration halo populations :
\begin{eqnarray}
    p^{\text{I}}_{NG} &=& 1 +\left( \frac{1}{1-\chi}\right)\frac{\partial\log\chi}{\partial\delta_{L}}\ ,\\ \label{p_low_mass_high_c}
    p^{\text{II}}_{NG} &=& 1 - \left(\frac{\eta}{1-\eta}\right)\left(\frac{\chi}{1-\chi}\right)\frac{\partial\log\chi}{\partial\delta_{L}}\ . \label{low_mass_low_c}
\end{eqnarray}
Note that since $\partial\log\chi/\partial\delta_{L} < 0$, $p^{\text{I}}_{NG} < 1$ and $p^{\text{II}}_{NG} > 1$. Our toy model thus predicts that high-concentration halos (at low mass) typically have a \textit{higher} value of $b_{\phi}$ than predicted by the universality relation, (i.e $p_{NG} < 1$) and lower concentration halos (at the same low mass) have a \textit{lower} value of $b_{\phi}$ than predicted by the universality relation (i.e $p_{NG} > 1$). This trend in the universality of halos is consistent with the results of our N-body simulations (see section~\ref{ssec:Results for Halos selected by mass and concentration}) as well as N-body simulations conducted by~\cite{Lazeyras:2022koc, Hadzhiyska:2025rez} and others.


\subsubsection{Intermediate mass halos selected by concentration}
\label{sssec:Intermediate mass halos selected by concentration}
The arguments of the preceding subsections~\ref{sssec:High mass halos selected by concentration} and~\ref{sssec:Low mass halos selected by concentration} indicate that the non-universal behaviour of halos selected by concentration is qualitatively different at the highest and lowest halo masses because halo concentration as an observable correlates with different lagrangian properties of their corresponding initial density peaks -- at the highest masses ($M\gg M_{*}$), halo concentration is correlated with peak curvature (leading to no non-universal behaviour) whereas at the lowest masses ($M\ll M_{*}$), halo concentration is correlated with the proximity of the initial halo peak to neighbouring massive peaks (leading to significant non-universal behaviour). The non-universality in the clustering of intermediate-mass halos selected by concentration is a combination of the two aforementioned effects. Obtaining an exact calculation of $p_{NG}$ for high- and low-concentration halos of intermediate mass from first principles (using the proposed model in subsection~\ref{ssec:Halos selected by mass and concentration}) is beyond the scope of this paper. However, we can nevertheless get a quantitative estimate for $p_{NG}$ for halos in different tertiles of concentration if we have a model for the concentration-mass probability density function (PDF). In fact, the concentration-mass PDF is nearly log-normal~\cite{ParanjapeSU} and is thus characterized by only two parameters, the mean and variance of the Gaussian PDF of log concentration at a given halo mass. Proceeding with this approach (following the calculations in~\cite{ParanjapeSU}), it is helpful to think in terms of a standardised log-concentration variable (defined for a given halo mass) :
\begin{eqnarray}
    s = \frac{\log(c/\overline{c}_{o})}{\sigma_{o}}\ ,\label{standardised log-concentration}
\end{eqnarray}
where $\log\overline{c}_{o}$ and $\sigma_{o}$ are the mean and standard deviation of the (log-normal) concentration-mass PDF in the \textit{global} universe (i.e in the absence of a background matter density $\delta_{L}$). The number density of halos of a certain mass and concentration can therefore be written as\footnote{In the expressions in this section we suppress the dependence on redshift whenever convenient.}~\cite{ParanjapeSU}:
\begin{eqnarray}
    n(M,c) = \overline{n}(M)\mathcal{P}(c|M)\ ,
\end{eqnarray}
where $\overline{n}$ is the number density of halos in the mass bin $M$ and $\mathcal{P}$ is the concentration-mass PDF: 
\begin{eqnarray}
    \mathcal{P}(c|M) = \frac{1}{c\sigma_{\log c}\sqrt{2\pi}}\exp\left(-\frac{\log(c/\overline{c})^{2}}{2\sigma^{2}_{o}}\right)\ .\label{log-normal}
\end{eqnarray}
where $\sigma_{\log c} $ and $\log \overline{c}$ are the standard deviation and the mean of log-concentration and are in general dependent on mass and redshift. In the absence of a background matter density perturbation, $\sigma_{\log c} = \sigma_{o}$ and $\log\overline{c} = \log\overline{c}_{o}$ respectively.

\indent Due to the log-normal nature of the concentration-mass PDF, $\overline{c}$ (in Eq.~\eqref{log-normal})  is also the \textit{median} concentration at a given halo mass. The dependence of $\overline{c}$ on the amplitude $A_{s}$ is given by the fitting formulas obtained by Diemer et.al.~\cite{Diemer:2018vmz,DiemerK}. On the other hand, the variance of the concentration-mass PDF $\sigma^{2}_{\log c}$ in a $\Lambda\text{CDM}$ cosmology is known to be largely independent of the peak-height and redshift~\cite{Bullock,Wechsler2002} (and $\sigma^{2}_{\log c} = \sigma^{2}_{o}\approx 0.15$~\cite{ParanjapeSU}) and therefore is independent of $A_{s}$ (since $A_{s}$ only affects the peak-height). We therefore have 
\begin{eqnarray}
    b_{\phi} &=& 2\frac{\partial\log n(M,c)}{\partial\log A_{s}}\ ,\nonumber\\
    &=&2 \frac{\partial\log\overline{n}}{\partial\log A_{s}} + 2\frac{\partial\log\mathcal{P}}{\partial\log\overline{c}}\cdot\left(\frac{\partial\log\overline{c}}{\partial\log A_{s}}\right)_{Diemer}\ ,\nonumber\\
    &=& \overline{b}_{\phi} + 2\frac{s}{\sigma_{o}}\frac{\partial\log\overline{c}}{\partial\log A_{s}}\ ,\label{bphi Diemer etal}
\end{eqnarray}
where $\overline{b}_{\phi} = 2\partial\log\overline{n}/\partial\log A_{s}$ and $\left(\partial\log\overline{c}/\partial\log A_{s}\right)$ is computed using the fitting formula given by Eq. (31) of~\cite{Diemer:2018vmz}. Note that $\overline{b}_{\phi}$ obeys the universality relation~\eqref{universality_mass_selected_halos} due to its sole dependence on the peak-height $\nu$.

\indent On the other hand, both the median concentration $\overline{c}$ and the scatter $\sigma_{o}$ change in the Separate Universe (i.e. in the presence of an ambient large-scale matter density). Their respective Separate Universe responses can be computed directly from Separate Universe simulations, as has been done in~\cite{ParanjapeSU}. ~\citet{ParanjapeSU} provide fitting formulas for responses of the mean $\mu(M|\delta_{L})$ and variance $\sigma^{2}_{s}(M|\delta_{L})$ of the standardised log-concentration $s$ (see Eq.~\eqref{standardised log-concentration}) up to second-order in the background density $\delta_{L}$ (see Table 1. of~\cite{ParanjapeSU}) for a flat $\Lambda\text{CDM}$ cosmology given by $\Omega_{m}=0.276$, $\Omega_{b}=0.045$, $h=0.7$, $n_s=0.961$ and $\sigma_{8}=0.811$. We therefore have,
\begin{eqnarray}
    \mu(M|\delta_{L},z) &=& \mu_{1}(M,z)\delta_{L} + \frac{1}{2}\mu_{2}(M,z)\delta^{2}_{L}+\dots\ ,\\
    \sigma^{2}_{s}(M|\delta_{L},z) &=& 1 + \Sigma_{1}(M,z)\delta_{L} + \frac{1}{2}\Sigma_{2}(M,z)\delta^{2}_{L} + \dots\ ,
\end{eqnarray}
with the result that 
\begin{eqnarray}
    \left(\frac{\partial\log\overline{c}}{\partial\delta_{L}}\right)_{\delta_{L}=0} &=& \sigma_{o}\mu_{1}(M,z)\ ,\\
    \left(\frac{\partial\log\sigma_{\log c}}{\partial\delta_{L}}\right)_{\delta_{L}=0} &=& \frac{1}{2}\Sigma_{1}(M,z)\ ,
\end{eqnarray}
where $\mu_{1}(M,z)$ and $\Sigma_{1}(M,z)$ are given by the fitting formulas derived in~\cite{ParanjapeSU} (Table 1.) for halo masses with peak-height $1.1\leq \nu\leq 2.9$ and $z<1.0$. The large-scale (lagrangian) bias of halos at a given mass and concentration is, therefore
\begin{eqnarray}
    b_{L}(M,c) &=& \frac{\partial\log\overline{n}}{\partial\delta_{L}} + \frac{\partial\log\mathcal{P}}{\partial\log\overline{c}}\cdot\frac{\partial\log\overline{c}}{\partial\delta_{L}}\nonumber\\
    &&+ \frac{\partial\log\mathcal{P}}{\partial\log\sigma_{\log c}}\cdot\frac{\partial\log\sigma_{\log c}}{\partial\delta_{L}}\ ,\\
    &=& \overline{b}_{L}(M) + \mu_{1}(M,z)s\nonumber\\
    &&+ \frac{1}{2}(s^{2}-1)\Sigma_{1}(M,z)\ ,\label{bL Paranjape etal}
\end{eqnarray}
where $\overline{b}_{L} = \partial\log\overline{n}(M,z)/\partial\delta_{L}$ is the lagrangian bias of halos selected purely by mass and obeys the universality relation~\eqref{universality_mass_selected_halos}.
From Eq.s~\eqref{bL Paranjape etal} and~\eqref{bphi Diemer etal}, we can compute the non-universality exhibited by halos of a given mass and concentration :
\begin{eqnarray}
    p_{NG}(M,c) &=& 1 + s\left(\mu_{1}-2\frac{R_{SU}}{\sigma_{o}}\frac{\partial\log\overline{c}}{\partial\log A_{s}}\right)\nonumber\\
    &&+ \frac{1}{2}(s^{2}-1)\Sigma_{1}\ .\label{pNG_Paranjape_Diemer}
\end{eqnarray}

\begin{figure}[h!]
    \centering
    \includegraphics[width=\columnwidth]{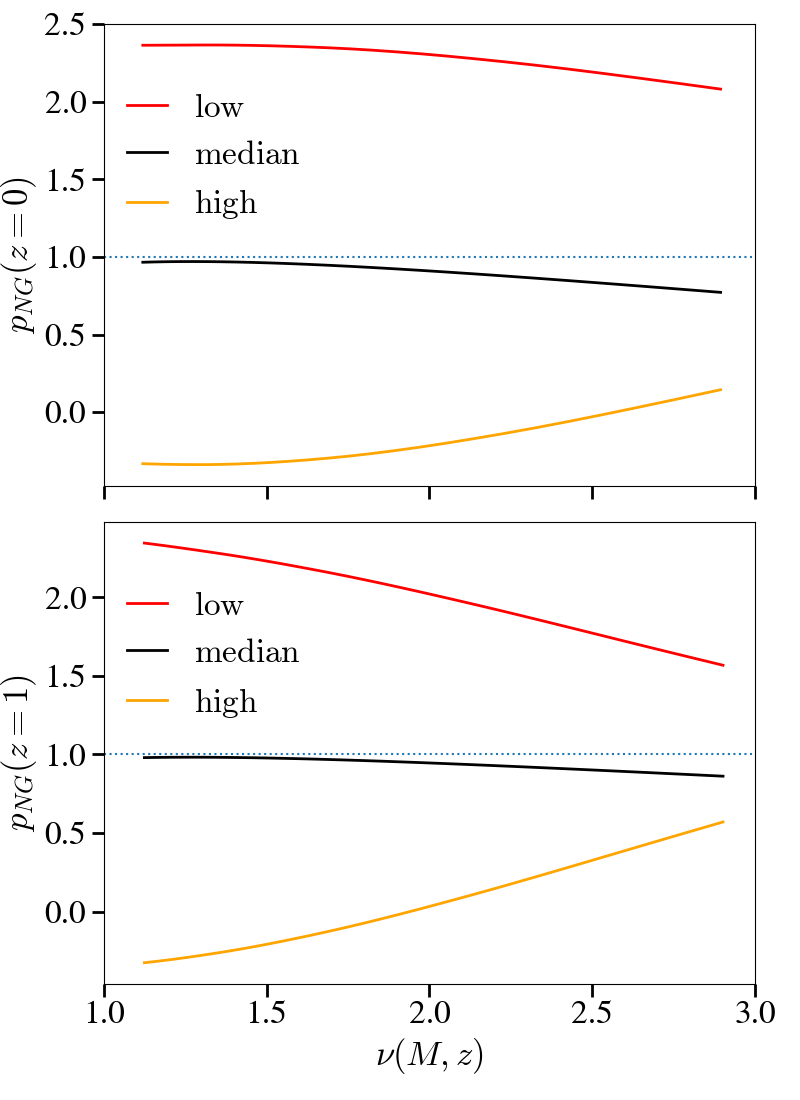}
    \caption{(Top) $p_{NG}$ for higher, median and lower tertiles of concentration computed using Eq.~\eqref{pNG_Paranjape_Diemer} for redshift $z=0$ (Bottom) $p_{NG}$ for higher, median and lower tertiles of concentration computed using Eq.~\eqref{pNG_Paranjape_Diemer} for redshift $z=1$. The dotted line represents $p_{NG}=1$. Halos in the median concentration tertile at both redshifts exhibit near-universality as $p_{NG}$ is very close to $1$ these halos.}
    \label{fig:pNG_Paranjape_Diemer}
\end{figure}

Figure~\ref{fig:pNG_Paranjape_Diemer} shows the values of $p_{NG}$ computed as a function of mass for different concentration tertiles using the above approach. For the plots in figure~\ref{fig:pNG_Paranjape_Diemer}, we use the same flat $\Lambda\text{CDM}$ cosmology as the one used by~\citet{ParanjapeSU} in computing the fitting formulas for $\mu_{1}(M,z)$ and $\Sigma_{1}(M,z)$. Note that halos in the median concentration tertile exhibit near-universal clustering with $p_{NG}\approx 1$ with higher (lower) concentration halos having lower (higher) $p_{NG}$. In fact, at $\nu\sim2.0$ (which is in the centre of the range of $\nu$ for which the fitting formulas in~\cite{ParanjapeSU} have been derived) $p_{NG}\approx 0.91$ at $z=0$ and $p_{NG}\approx 0.95$ at $z=1$. Similarly for the higher concentration tertile, $p_{NG}\approx 2.31$($z=0$) and $p_{NG}\approx 2.02$($z=1$) and for the lower halo concentration tertile, $p_{NG}\approx -0.22$ ($z=0$) and $p_{NG}\approx 0.03$ ($z=1$). Similarly the degree of non-universality (in terms of $|p_{NG}-1|$) is lower for high-mass halos ($\nu\gtrsim 2.5$) at $z=1$ in consistency with the fact that halos of this mass are effectively halos with high peak-height that grow in a more matter-dominated universe (see subsection~\ref{sssec:High mass halos selected by concentration}). As we shall see in section~\ref{sec:Results}, calculations using the approach outlined in this subsection (i.e. with the formula~\eqref{pNG_Paranjape_Diemer}) do indeed capture the trend in the non-universality exhibited by concentration-selected halos in our $\Lambda\text{CDM}$ N-body simulations (see figure~\ref{fig:LCDM_bphi_vs_b0_all_nu}).  
\section{Methods}
\label{sec:Methods}
In this section, we describe the details of the N-body simulations performed for this paper and the methodology for identifying dark matter halos and computing halo concentrations, halo bias $b_{h}$ and $b_{\phi}$ for halos in different mass and concentration bins.

\subsection{N-body simulations}
\label{ssec:N-body simulations}

All N-body simulations were performed using the Gadget-4 TreePM code~\cite{Gadget4}. Initial conditions for each simulation were generated at redshift $z=49$ (corresponding to $a=0.02$) using the second order lagrangian perturbation theory (2LPT) initial conditions generator in Gadget4~\cite{Gadget4}. We performed two sets of simulations -- one with a scale-free cosmology (pure EdS with power law initial power spectrum $P_{lin}(k) \propto k^{-2}$) and one with a flat, $\Lambda\text{CDM}$ cosmology. 

Scale-free N-body simulations with a power-law initial power spectrum present a way to probe halo clustering properties over a large range of halo masses with relatively low computational cost. This is because due to the absence of any natural mass-scale in such cosmologies, halo clustering properties depend on the halo mass $M$ and redshift $z$ only through the ratio $M/M_{*}(z)$ where $M_{*}(z)$ is the non-linear mass at redshift $z$ defined by the relation $\nu(M_{*}(z),z) = 1.0$. The non-linear mass-scale decreases rapidly with increasing redshift (going as $a^{6}$ for $P(k)\propto k^{-2}$). Hence considering outputs of scale-free simulations at different redshifts at relatively smaller box-sizes and particle mass resolutions enables one to effectively probe a large range of halo masses (through a large range in $M/M_{*}$). For the scale-free simulations performed for this paper, we chose a box of length $L_{\text{box}} = 400\ \text{Mpc/h}$ and $N_{p} = (720)^{3}$ number of particles. We choose an initial power spectrum that follows a power-law $P(k)\propto k^{-2}$ with $\sigma_{8}=0.75$. We also run pairs of high-$\sigma_{8}$ and low-$\sigma_{8}$ with $\sigma_{8}=0.75\pm 0.02$ and with all other parameters (including the seed for generation of initial conditions) unchanged, in order to measure $b_{\phi}$(see subsection~\ref{ssec:Measurement of b and bphi}). We adopt a softening length of $35\ \text{kpc/h}$ and obtain five output snapshots at times evenly spaced in $\log a$ from $a=0.4$ through $a=1$. This choice of simulation parameters allows us to reliably probe halo clustering properties in different tertiles of halo concentration for halo masses corresponding to peak-heights $1.1\leq\nu\leq 3.0$ (see subsection~\ref{ssec: Halo finding and concentration measurements}). Since each snapshot probes a different range of peak-heights, we combine the measurements of $b_{L}$ and $b_{\phi}$ from all five snapshots and use a smoothing spline to estimate the functional dependences $b_{L}(\nu)$ and $b_{\phi}(\nu)$ in each tertile of concentration.  

For the $\Lambda\text{CDM}$ dark-matter only simulations, we choose a box length $L_{\text{box}}=490\ \text{Mpc/h}$ and $N_{p}=(640)^{3}$ number of particles with cosmological parameters $\Omega_{cdm}=0.25$, $\Omega_{b}=0.05$, $h=0.7$, $\sigma_{8}=0.83$ and $n_{s}=0.95$. For the purposes of this work, we regard CDM and baryons as a single `matter' fluid with $\Omega_{m} = 0.30$.  To generate initial conditions for this set of simulations, we provide an input power spectrum computed for our chosen cosmology\footnote{Note that this cosmology is different from the one used to calculate the $b_{\phi}-b_{h}$ relation in subsection~\ref{sssec:Intermediate mass halos selected by concentration}.} using the publicly available boltzmann solver CLASS~\cite{Lesgourgues:2011re}. Similar to the scale-free simulations, we also perform high-$\sigma_{8}$ and low-$\sigma_{8}$ simulations with $\sigma_{8}\pm 0.02$ (and all other parameters as well as the seed used to generate initial conditions unchanged) in order to compute $b_{\phi}$ (see subsection~\ref{ssec:Measurement of b and bphi}). For the $\Lambda\text{CDM}$ simulations, we adopt a softening length of $50\ \text{Mpc/h}$ and measure halo bias $b_{h}$ and $b_{\phi}$ for three output times evenly spaced in $\log a$ between $z=1$ and $z=0$. This corresponds to output redshifts (scale-factors) $z=1 (a=0.5)$, $z=0.414 (a=0.707)$, and $z=0 (a=1)$. Our choice of parameters and mass resolution allows us to probe halo clustering properties in different tertiles of halo concentration for halos mass $M\geq 1.5\times 10^{13}\ M_{\odot}/h$ (see subsection~\ref{ssec: Halo finding and concentration measurements}).

\subsection{Halo-finding and concentration measurements}
\label{ssec: Halo finding and concentration measurements}

We identify halos in our N-body simulations using the phase-space halo finder ROCKSTAR~\cite{RockstarPaper} which identifies halos using the Friends-of-Friends algorithm and assigns masses using a Spherical Overdensity (SO) mass-assignment scheme. For the SO mass-assignment, we use the virial density threshold which in general can be redshift-dependent~\cite{BryanNorman}. While we restrict our halo catalogs to halos containing no fewer than 100 particles, we perform all halo concentration and bias/$b_{\phi}$ measurements for halos with at least 400 particles. This corresponds to a minimum halo mass of $M_{\text{min}} = 2\times 10^{13}\ M_{\odot}/h$ and $M_{\text{min}} = 1.5\times 10^{13}\ M_{\odot}/h$ for the scale-free and $\Lambda\text{CDM}$ simulations respectively. Moreover, we only consider halos with the virial ratio $T/|U|\leq 1$ -- thus ensuring that they are sufficiently relaxed.

For the purposes of analysing the dependence of halo clustering properties $b_{h}$ and $b_{\phi}$ on halo concentration, we split our halo catalogue into tertiles of the halo concentration. We estimate halo concentration $c_{vir} = R_{vir}/R_{s}$ from the maximum circular velocity $V_{max}$ within the halo, halo mass $M_{vir}$ and halo radius $R_{vir}$ by inverting the following relation~\cite{RockstarPaper,Klypin:2010qw} :
\begin{eqnarray}
    \frac{c_{vir}}{f(c_{vir})} = V^{2}_{max}\frac{R_{vir}}{GM_{vir}}\frac{2.1626}{f(2.1626)}\ ,
    \label{Klypin_concentrations}
\end{eqnarray}
where
\begin{eqnarray}
    f(x) = \log (1+x) -\frac{x}{1+x}\ .
\end{eqnarray}
The above method of computing halo concentrations is less sensitive (than directly taking the ratio $R_{vir}/R_{s}$) to uncertainties/biases in the fitting of halo density profiles to the NFW form which may arise from limitations of particle mass resolution~\cite{Klypin:2010qw,RockstarPaper}. We find that virial concentrations found using Eq.~\eqref{Klypin_concentrations} systematically differ from virial concentrations estimated directly by calculating the ratio $R_{vir}/R_{s}$; but the two estimates of concentration are very tightly correlated. In particular we find a Spearman correlation of $r\gtrsim 0.8$ between the two estimates of concentration for the mass range of interest, suggesting a strong monotonic relation between them. This indicates that results obtained for different tertiles of halo concentration are not significantly dependent on the choice of the concentration estimate.

\subsection{Measurement of halo bias and $b_{\phi}$}
\label{ssec:Measurement of b and bphi}

  We compute the large-scale (Eulerian) halo bias $b_{h}$ directly by estimating the following limit of the ratio of the matter-halo power spectrum ($P_{mh}$) to the matter-matter power spectrum ($P_{mm}$) :
\begin{eqnarray}
    b_{h}(\geq M,z) = \lim_{k\rightarrow 0} \frac{P_{mh}(k,z)}{P_{mm}(k,z)}\ .
\end{eqnarray}
To compute matter-halo and matter-matter power spectrum, we construct the matter-density and halo number density fields using a cloud-in-cell mass assignment and then fourier-transform them using FFTW3~\cite{Frigo:2005zln}. To compute the limit as $k\rightarrow 0$, we fit a polynomial of the form $b_{h} = b_{0}+b_{2}k^{2}$ to the ratio $P_{mh}/P_{mm}$ up to $k_{max}=0.15\ \text{h/Mpc}$ (see~\cite{Lazeyras:2017hxw}). 

The above method gives reliable estimates of halo bias (and their uncertainties) for all but the highest bias objects observed in our simulations. For high-bias objects, the fitting formula $b_{h} = b_{0} + b_{2}k^{2}$ may not adequately capture all the non-linear contributions to the ratio $P_{mh}/P_{mm}$ -- because such objects cluster non-linearly over most of the range $0 < k < 0.15\ \text{h/Mpc}$ used for the bias fit. For example, note that for the scale-free simulation, the matter clustering amplitude $\Delta_{m}$ at scales $k\sim 0.15\ \text{h/Mpc}$ and redshift corresponding to $a=0.4$ is  $ \Delta_{m}(k) = \sqrt{k^{3}P_{mm}(k,a=0.4)/2\pi^2} \sim 0.24$ whereas the clustering amplitude for the highest-bias objects ($b_{L}\sim 8$) observed in the scale-free simulations at the same scale and redshift is $\Delta_{h} = b_{h}\Delta_{m} \gtrsim 2 $ -- meaning that objects (in the scale-free simulations) with bias $b_{L} \gtrsim 8$ cluster non-linearly over most of the range $0 < k < 0.15\ \text{h/Mpc}$ used for the bias fit. This fact can lead to poorly controlled uncertainties in the estimation of halo bias of high-bias objects observed in our simulations.  

  We compute $b_{\phi}$ from the difference in the abundances of halos in the high-$\sigma_{8}$ and low-$\sigma_{8}$ simulations as follows :
  \begin{eqnarray}
      b_{\phi}(\geq M, z) = 2\frac{\partial\log n_{h}}{\partial\log A_{s}} = \frac{\sigma_{8}}{n_{h}}\cdot\left(\frac{n^{+}_{h}-n^{+}_{h}}{\sigma^{+}_{8}-\sigma^{-}_{8}}\right)\ ,
  \end{eqnarray}
  where $\sigma^{\pm}_{8}$ and $n^{\pm}_{h}$ represent the mass variance and the number density of halos in the high-$\sigma_{8}$ and low-$\sigma_{8}$ simulations respectively; while $\sigma_{8}$ and $n_{h}$ represent the same quantities measured in the `fiducial' simulations (i.e. the scale-free simulations with $\sigma_{8}=0.75$ and $\Lambda\text{CDM}$ simulations with $\sigma_{8} = 0.83$).  

  Proceeding further, we obtain more reliable estimates of the halo bias $b_{h}$, $b_{\phi}$ (with errorbars) by bootstrapping over 33 different realisations of the initial conditions in the case of scale-free simulations and 24 different realisations of the initial conditions in the case of $\Lambda\text{CDM}$. In the same way, we also obtain reliable estimates of the ratio $b_{\phi}/b_{L}\delta_{c}$ (with errorbars) for each tertile of concentration and for all halo peak-heights probed by our scale-free simulations -- this enables us to better understand the behaviour of the $b_{\phi}-b_{h}$ relationship as a function of halo peak-height in the scale-free cosmology considered here. 
  
  For each realisation of the initial conditions, we perform three simulations -- a `fiducial' simulation, a high-$\sigma_{8}$ simulation and a low-$\sigma_{8}$ simulation and compute estimators of the halo bias and $b_{\phi}$ using the methods described above. Moreover, we compute halo bias $b_{h}$ and $b_{\phi}$ (as well as the ratio $b_{\phi}/b_{L}\delta_{c}$) in \textit{cumulative} mass bins -  i.e. $b_{h} = b_{h}(\geq{M})$ and $b_{\phi} = b_{\phi}(\geq M)$. 
  
  The estimates of $b_{\phi}$ (and of the errorbars) obtained using the method described above are pretty sensitive to halo shot noise, especially for the rarer, high bias halos. This is because, unlike the measurement of $b_{h}$, measurement of $b_{\phi}$ proceeds with directly subtracting halo abundances in the high-$\sigma_{8}$ and the low-$\sigma_{8}$ simulations. While we cannot completely eliminate this uncertainty, we mitigate its effects by quoting our results for halo masses $M$ which satisfy the requirement that there be at least $50$ halos more massive than $M$ in a single simulation box and/or in each concentration tertile (for halos selected by mass and concentration). 

\section{Results}
\label{sec:Results}
\subsection{Halos selected by mass}
\label{ssec:Results for Halos selected by mass}

\begin{figure}[!hpt]
    \centering
    \includegraphics[width=\columnwidth]{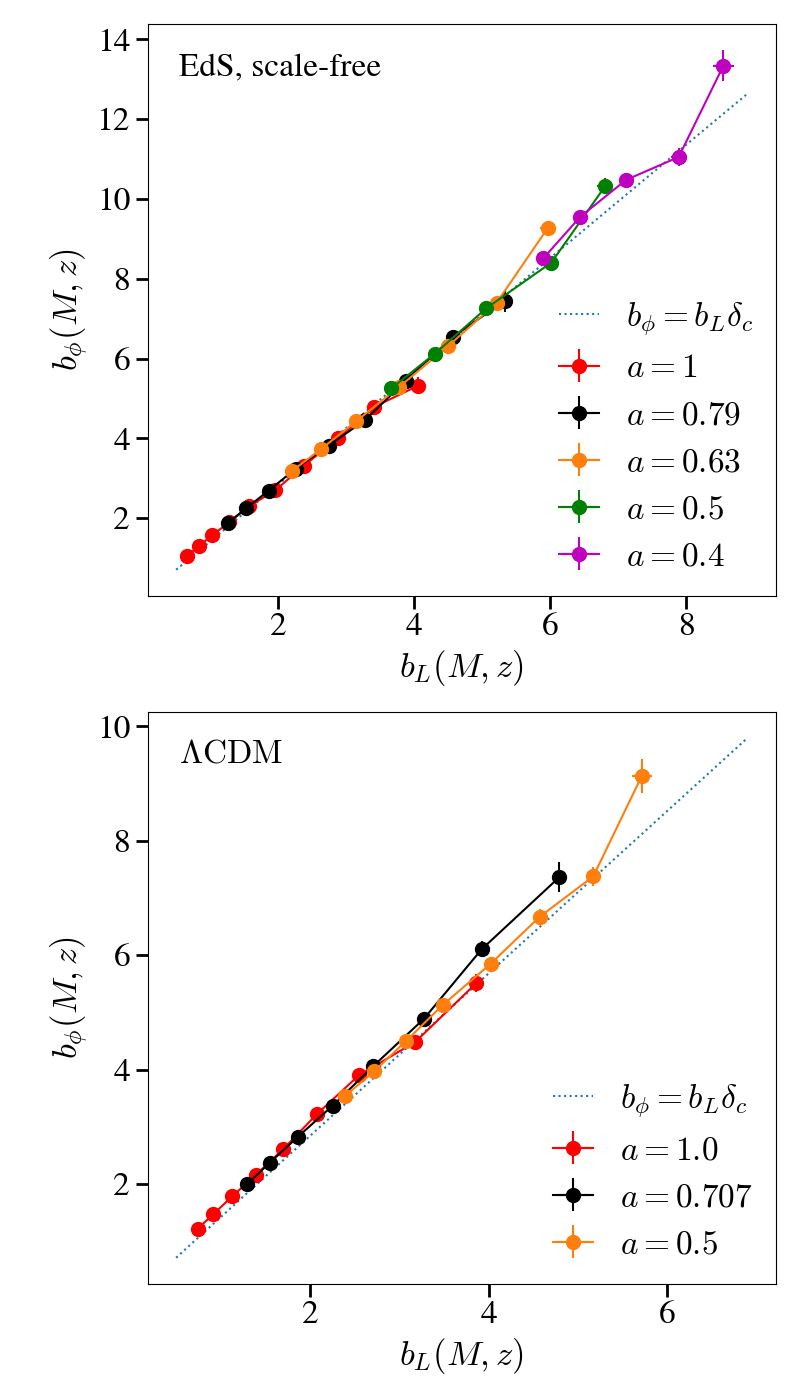}
    \caption{(Top) Plot of $b_{\phi}$ and lagrangian halo bias $b_{L}$ for halos in the scale-free simulations selected only by their virial mass at snapshots with scale factors evenly spread in $\log a$ between $a=1$ (i.e. $z=0$) and $a=0.5$ ($z=1.0$).(Bottom) Plot of $b_{\phi}$ and lagrangian halo bias $b_{L}$ for halos in the scale-free simulations selected only by their virial mass at snapshots with scale factors evenly spread in $\log a$ between $a=1$ (i.e. $z=0$) and $a=0.4$ ($z=1.5$). The dotted line shows the relation between $b_{\phi}$ and $b_{L}$ expected from universality (see equation~\eqref{universality_def}). }
    \label{fig:EdS_scale_free_LCDM_bphi_vs_bL_Mvir}
\end{figure}



Figure~\ref{fig:EdS_scale_free_LCDM_bphi_vs_bL_Mvir} (top and bottom) shows the relation between the lagrangian halo bias $b_{L}$ and $b_{\phi}$ for halos selected only by their virial mass observed in the scale-free (top) and $\Lambda\text{CDM}$ (bottom) simulations respectively. We see that for both simulation sets, the observed $b_{\phi}-b_{L}$ relation very closely follows the trend expected from the universality relation~\eqref{universality_mass_selected_halos} albeit with $R^{-1}_{SU}\sim\delta_{c}\approx 1.42$ (as might be expected for FOF halos with SO masses -- see discussion at the end of subsection~\ref{ssec:Halos selected by mass}).

Note that the measured $b_{\phi}-b_{L}$ relation shows a slight disagreement with the universality relation at the highest halo biases and earliest times (particularly the highest bias data point in both figures) -- at these data points, the measured $b_{\phi}-b_{L}$ for both scale-free and $\Lambda\text{CDM}$ simulations seems to follow $b_{\phi}\approx b_{L}\times 1.6$ rather than $b_{\phi}\approx b_{L}\times 1.42$. We attribute this to poorly controlled uncertainties in estimation of $b_{\phi}$ as well as in the measurement of the halo bias of high-bias objects (see subsection~\ref{ssec:Measurement of b and bphi}).

\subsection{Halos selected by mass and concentration}
\label{ssec:Results for Halos selected by mass and concentration}

\begin{figure}[!hpt]
    \centering
    \includegraphics[width=\columnwidth]{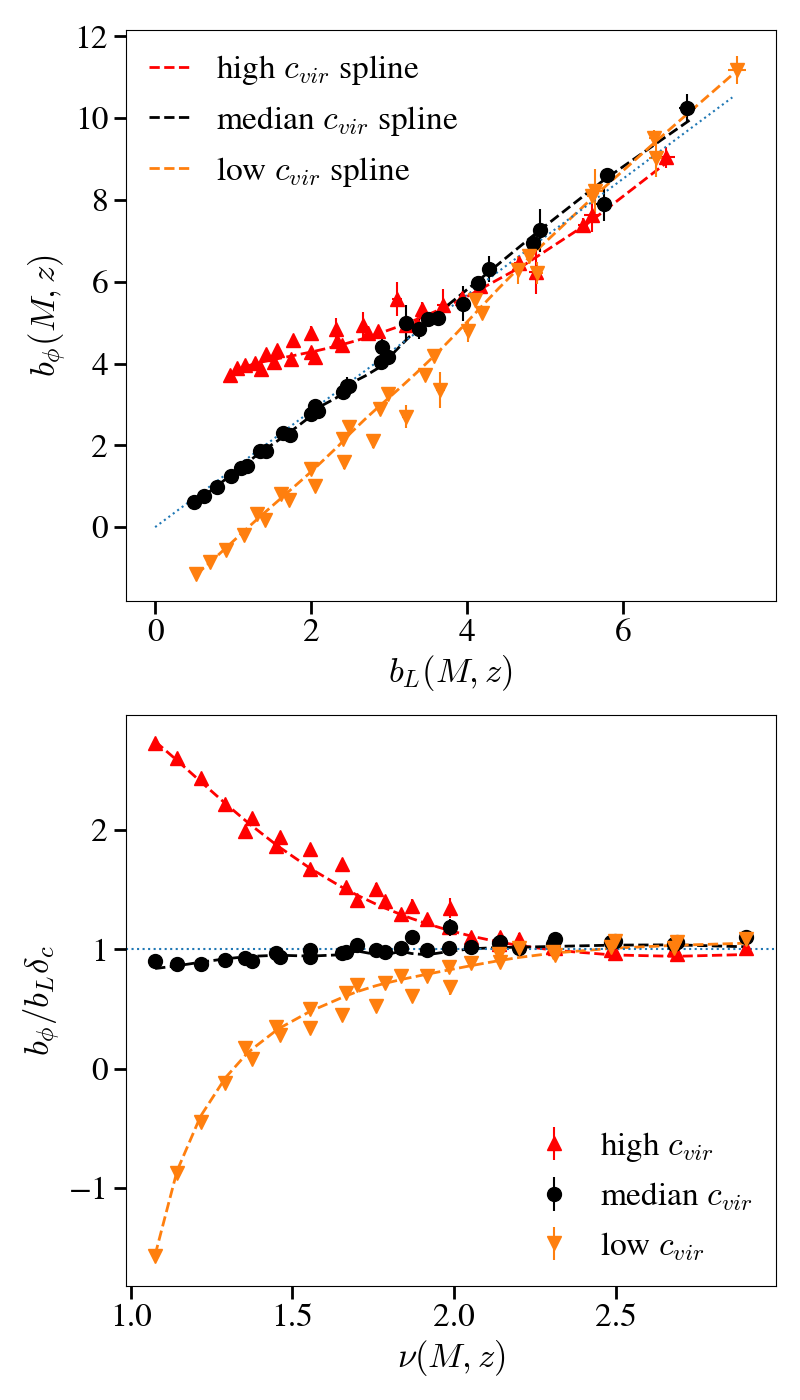}
    \caption{(Top) Plot of $b_{\phi}$ and lagrangian halo bias $b_{L}$ in different tertiles of concentration for all halo peak-heights measured in our scale-free-simulations. (Bottom) Plot of the ratio $b_{\phi}/b_{L}\delta_{c}$ in different tertiles of concentration measured for all halo peak-heights in our scale-free simulations. In both the plots, the blue dotted line shows the predictions from the universality relation~\eqref{universality_def} and the dashed lines are obtained from the spline-interpolated functional forms $b_{L}(\nu)$ and $b_{\phi}(\nu)$ obtained for all halo peak-heights (and in each tertile of concentration) measured in our scale-free simulations}
    \label{fig:EdS_bphi_vs_b0_all_nu}
\end{figure}

\begin{figure}[!hpt]
    \centering
    \includegraphics[width=\columnwidth]{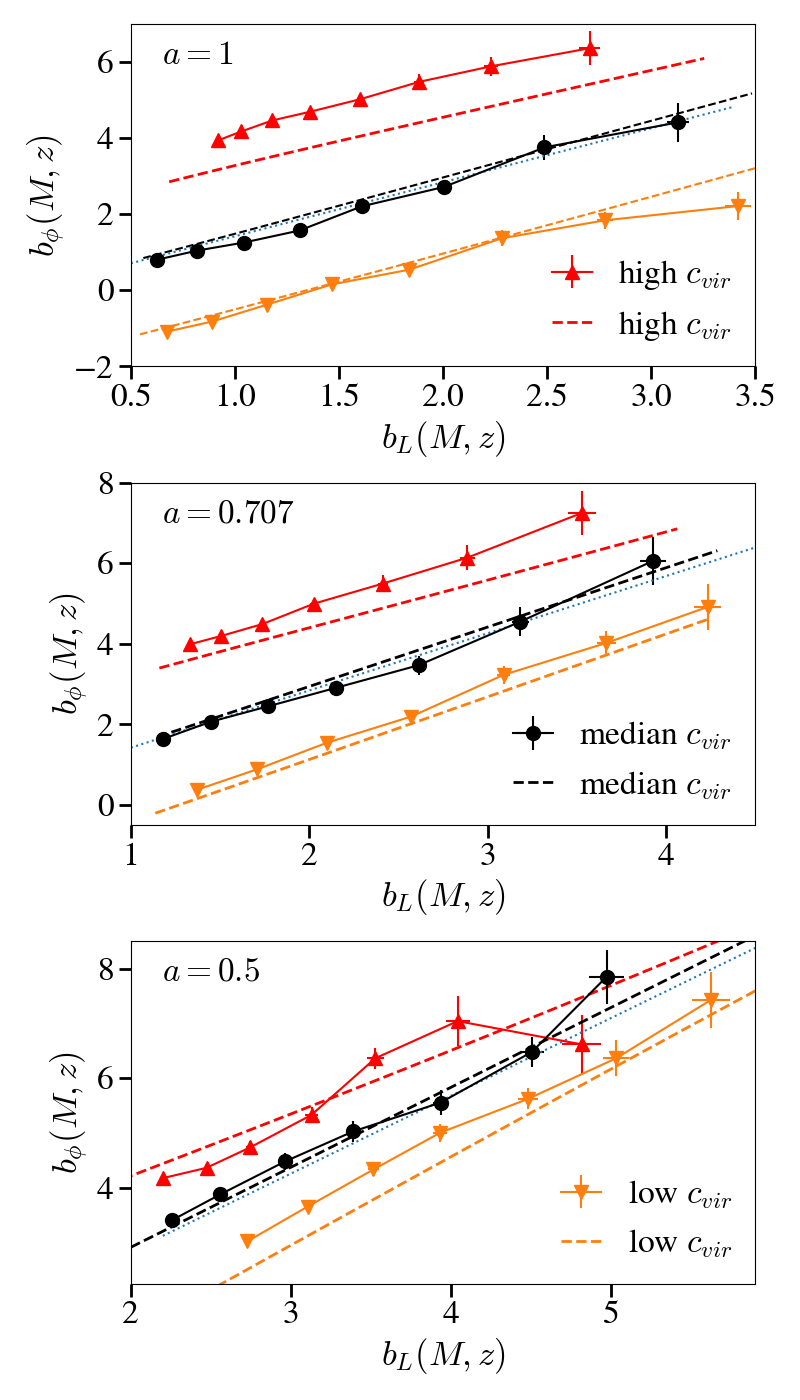}
    \caption{Plot of $b_{\phi}$ and $b_{L}$ for halos in different tertiles of concentration at different redshifts measured in our $\Lambda\text{CDM}$ simulations. The dashed lines represent the theoretical predictions computed from Eq.s~\eqref{bL Paranjape etal} and~\eqref{bphi Diemer etal} using the results of~\cite{ParanjapeSU} The blue dotted line represents the universality relation~\eqref{universality_def}.}
    \label{fig:LCDM_bphi_vs_b0_all_nu}
\end{figure}


Figure~\ref{fig:EdS_bphi_vs_b0_all_nu} (top) shows the relation between $b_{\phi}$ and $b_{L}$ in each tertile of concentration and for all halo peak-heights observed in our scale-free simulations. For the scale-free simulations,  Figure~\ref{fig:EdS_bphi_vs_b0_all_nu} (bottom) shows the deviation from universality parameterised by the ratio, $b_{\phi}/b_{L}\delta_{c}$ for halos in each tertile of concentration and for all peak-heights measured in the scale-free simulations. Figure~\ref{fig:EdS_bphi_vs_b0_all_nu} (top) shows that for large-bias objects ($b_{L}\gtrsim 4$), the $b_{\phi}-b_{L}$ relationships approaches the universality relation (Eq.~\eqref{universality_def}) while for lower halo biases, higher (lower) concentration halos exhibit higher(lower) $b_{\phi}$ values compared to the predictions of the universality relation. On the other hand, halos in the median concentration tertile show nearly universal behaviour even for lower halo biases. Note that the low-bias behaviour seen here corroborates the qualitative predictions of subsection~\ref{sssec:Low mass halos selected by concentration} which argues that non-universality in low-mass, concentration-selected halos arises as a result of halo mass accretion rate being affected by the gravitational effect of neighbouring, higher mass halos.

Going further, figure~\ref{fig:EdS_bphi_vs_b0_all_nu} (bottom) shows that the ratio $b_{\phi}/b_{L}\delta_{c}$ approaches unity at large peak-heights. In fact, the ratio $b_{\phi}/b_{L}\delta_{c}$ is within $\lesssim 5\%$ of unity for higher peak-heights ($\nu\gtrsim 2.25$) and for \textit{all} tertiles of concentration -- indicating that $b_{\phi}\approx b_{L}\delta_{c}$ (within $5\%$) for all concentration tertiles. In terms of the Separate Universe picture (see subsection~\ref{sec:Halo bias and the Separate Universe} and equations~\eqref{SU_bphi_expression} and~\eqref{SU_bias_expression}), this indicates that for concentration-selected halos in this regime, the contribution of small-scale statistics $\theta_{i}$ which are \textit{non-universal} (i.e. $d\theta_{i}/d\delta_{L}\cdot\delta_{c}\neq 2\cdot d\theta_{i}/d\log A_{s}$) to the lagrangian bias (through Eq.~\eqref{SU_bias_expression}) is $\lesssim 5\%$.     

Our results from scale-free simulations support, with high statistical significance, the hypothesis that high-mass halos ($\nu\gtrsim 2.25$) in scale-free simulations selected by mass and concentration do not show any non-universality in their clustering (unlike their low-mass counterparts). They also corroborate the explanation in subsection~\ref{sssec:High mass halos selected by concentration} whereby high-mass halos selected by concentration obey the universality relation due to their concentrations being most tightly correlated to their respective peak curvatures. 

Figure~\ref{fig:LCDM_bphi_vs_b0_all_nu} shows the $b_{\phi}-b_{L}$ relationships in each tertile of concentration for halo masses observed out to redshift $z=1$ in our $\Lambda\text{CDM}$ simulations. The dashed lines in figure~\ref{fig:LCDM_bphi_vs_b0_all_nu} show the calculation of $b_{\phi}$ and $b_{L}$ using equations~\eqref{bL Paranjape etal} and~\eqref{bphi Diemer etal}. These equations in turn use the Separate Universe calculations of~\cite{ParanjapeSU} and the concentration-mass relation obtained in~\cite{Diemer:2018vmz}. Figure~\ref{fig:LCDM_bphi_vs_b0_all_nu} illustrates that while these theoretical computations may not fully reproduce the $b_{\phi}-b_{L}$ relation observed in our simulations, they nevertheless capture the correct trend of decreasing non-universality with redshift\footnote{Note that these calculations have really been done for the $\Lambda\text{CDM}$ parameters in ~\cite{ParanjapeSU}, which is different from the one assumed for our $\Lambda\text{CDM}$ simulations. See subsection~\ref{sssec:Intermediate mass halos selected by concentration} for more information.}. Although our $\Lambda\text{CDM}$ simulations do not probe halo masses high enough to capture the onset of universality (and the fitting formulas obtained by~\citet{ParanjapeSU} do not extend to redshifts higher than $z=1$), we can still see from figure~\ref{fig:LCDM_bphi_vs_b0_all_nu} that the extent of non-universality of concentration-selected halos steadily decreases for high-redshift and high-bias halos as might be expected for the fact that such halos are high-mass ($M\gg M_{*}$) halos which form in a matter-dominated universe (see section~\ref{sssec:High mass halos selected by concentration}.    

As commented in subsection~\ref{sssec:High mass halos selected by concentration}, high redshift halos in a $\Lambda\text{CDM}$ cosmology really formed in a matter-dominated cosmology and would be expected to show universality for high halo masses (or in other words, for high halo peak-heights). The threshold peak-height beyond which universality is restored ($\nu\gtrsim 2.25 $) corresponds to a halo mass of $M_{h}\sim 10^{12}-10^{13}\ M_{\odot}/h$ at redshifts $z\sim 3$ in typical $\Lambda\text{CDM}$ cosmologies. This falls within the range of quasar/Lyman Break Galaxy (LBG) halo masses~\cite{Kim:2007de, Ishikawa_2017} at redshifts $z\sim 3$ -- indicating that the non-universality observed for high redshift quasars or LBGs is likely an artifact of the selection function thereof and not due to non-universal clustering of the underlying halos.

\section{Discussion}
\label{sec:Discussion}

The quasi-local nature of halo formation makes it possible to understand the connection between halo bias $b_{h}$ and the non-Gaussian bias parameter $b_{\phi}$ in terms of the responses of relevant small-scale statistics $\{\theta_{i}\}$ (constructed from matter fluctuation fields) to ambient large-scale matter density and primordial curvature fluctuations respectively (see section~\ref{sec:Halo bias and the Separate Universe}). The response to an ambient large-scale matter density can be computed using Separate Universe simulations whereas the response to an increase in the amplitude of primordial curvature fluctuations can be computed from simulations by appropriately varying the amplitude of the primordial power spectrum (i.e. by varying $A_{s}$).   Universality in this picture is the special case where both responses (to large-scale matter density and to curvature fluctuations) are proportional to each other according to Eq.~\eqref{universality_statistics} for all the relevant statistics $\{\theta_{i}\}$ that govern halo abundance. In this paper, we use this approach to understand non-universality in the clustering of halos selected by mass and/or concentration.

For halos selected purely by mass, the only relevant statistic that governs their abundance is the mass variance $\sigma_{R}$ which is the matter power spectrum smoothed on lagrangian scale $R$ corresponding to the relevant halo mass. In section~\ref{sec:Halo bias and the Separate Universe}, subsection~\ref{ssec:Halos selected by mass}, we argue that the universality observed in the clustering of purely mass-selected halos follows from the fact that the lagrangian bias of such halos can be expressed as (see subsection~\ref{ssec:Halos selected by mass}) : 
\begin{eqnarray}
    b_{L}(M) &=& \frac{d\log n_{h}(M)}{d\delta_{L}}\nonumber\ ,\\
             &=& 2\cdot\frac{\partial\log n_{h}}{\partial\log A_{s}}\cdot R_{SU}\nonumber\ ,\\
             &\approx& b_{\phi}(M)\delta_{c}^{-1}\ ,
\end{eqnarray}
where $R_{SU} = d\log\sigma_{R}/d\delta_L$ is the response of $\sigma_{R}$ to an ambient large-scale matter density fluctuation $\delta_{L}$ computed within the Separate Universe framework. The results of our N-body simulations (for both scale-free cosmology as well as $\Lambda\text{CDM}$ cosmology, see subsection~\ref{ssec:Results for Halos selected by mass}) indeed show that halos selected by mass alone obey the universality relation (see figure~\ref{fig:EdS_scale_free_LCDM_bphi_vs_bL_Mvir}), albeit with $\delta_{c}\approx 1.42$ as is known to hold for the use of FOF halo-finding algorithm together with SO mass-assignment schemes for determining halo mass (see discussion at the end of subsection~\ref{ssec:Halos selected by mass}). In this way, the Separate Universe framework provides a natural explanation for the observed universality in the clustering of dark matter halos selected only by mass.

In section~\ref{sec:Halo bias and the Separate Universe}, subsection~\ref{ssec:Halos selected by mass and concentration}, we propose a qualitative/semi-quantitative explanation for the non-universality in the clustering of halos selected by mass and concentration. Our explanation in subsection~\ref{ssec:Halos selected by mass and concentration} proceeds along similar lines as the explanation (proposed in~\cite{TidalEffects,TidalEffects2,Dalal:2008zd,Ramakrishnan:2019wtt} for the inversion in the trend of halo assembly bias w.r.t concentration seen at low ($M\ll M_{*}$) halo masses and is primarily based on the observation that concentration is largely correlated with different properties of the halo progenitors at high ($M\gg M_{*}$) and low ($M\ll M_{*}$) halo masses~\cite{Dalal:2008zd,Ramakrishnan:2019wtt,TidalEffects,TidalEffects2}. High-mass ($M\gg M_{*}$) halos form from the collapse of large, rare peaks in the primordial matter-density field. The concentration of such halos is largely determined by lagrangian properties of their progenitor peak -- in particular by the peak curvature $\xi_R = R^2\nabla^{2}\delta_R$ where $R$ is the lagrangian scale associated with the relevant halo mass. In scale-free, purely matter dominated cosmologies, the abundance of high-mass ($M\gg M_{*}$) halos segregated by concentration is primarily governed not only by $\sigma_{R}$ but also by the variance of the peak curvature $\sigma_{\xi,R}^{2}$ and the correlation coefficient $\gamma_R = \langle\delta_{R}\xi_{R}\rangle/\sigma_{R}\sigma_{\xi,R}$. Both $\sigma_R$ and $\sigma_{\xi,R}$ vary in the same way under an ambient large-scale density perturbation as well as in response to an increase in the amplitude of primordial fluctuations. On the other hand, the correlation coefficent $\gamma_R$ is unchanged in both cases (see Appendix ~\ref{app:responses}). As a result the bias and $b_{\phi}$ of high-mass, concentration-selected halos in purely matter-dominated universes obey the universality relation (see subsection~\ref{sssec:High mass halos selected by concentration}) 
\begin{eqnarray}
    b_{\phi} = (b_{h}-1)R_{SU}^{-1} \approx (b_{h}-1)\delta_{c}\ .
\end{eqnarray}

The results of our scale-free simulations (see section~\ref{ssec:Results for Halos selected by mass and concentration}) are consistent with this conclusion. In particular, figure~\ref{fig:EdS_bphi_vs_b0_all_nu} shows that for halos with high peak-heights ($\nu\gtrsim 2.25$), the ratio $b_{\phi}/((b_{h}-1)\delta_c)$ is within $5\%$ of unity for halos in all tertiles of concentration. For typical $\Lambda\text{CDM}$ cosmologies, the median concentration of halos also depends on the scale-independent linear growth rate $f(z) = d\log D(z)/d\log a$ which is significantly different from unity at late times and in principle would contribute to the non-universality in mass and concentration-selected halos. However, as we see in the results of our $\Lambda\text{CDM}$ simulations (figure~\ref{fig:LCDM_bphi_vs_b0_all_nu}), the extent of non-universality observed in halos selected by mass and concentration decreases with increasing redshift as would be expected from the fact that the universe becomes more and more matter dominated at high redshifts and the non-linear mass scale decreases rapidly with increasing redshifts (so that halos of a given mass approach the high-mass ($M\gg M_{*}$) where their clustering becomes universal).   

On the other hand, low-mass ($M\ll M_{*}$) halos form from the collapse of smaller, more abundant peaks in the primordial matter density field. The mass accretion rate of such peaks is very strongly influenced by tidal stripping caused by the gravitational influence of a neighbouring massive peak~\cite{Dalal:2008zd,TidalEffects,TidalEffects2}. In the low-mass limit, the concentration of halos is therefore most tightly correlated not with the lagrangian properties of their progenitor peaks but with properties of their immediate, small-scale environment -- especially the tidal shear caused by the gravitational effect of neighbouring high-mass halos~\cite{Dalal:2008zd,Ramakrishnan:2019wtt,TidalEffects,TidalEffects2}.
In subsection~\ref{sssec:Low mass halos selected by concentration} we present a heuristic model based on the aforementioned connection of concentration with halo environment at low masses (built along the lines of the toy model proposed by~\citet{Dalal:2008zd} to explain assembly bias trends w.r.t concentration) which, under reasonable assumptions, can qualitatively explain the non-universality observed in low-mass ($M\ll M_{*}$) halos segregated by concentration. Our toy model correctly predicts that, at low halo masses, high (low) concentration halos have a higher (lower) value of $b_{\phi}$ than predicted by the universality relation. In other words, the $b_{\phi}$ and halo bias $b_{h}$ of concentration-selected halos at low mass obey a relation of the kind $b_{\phi} = (b_{h}-p_{NG})\delta_c$ with $p_{NG}<1$ for high-concentration halos and $p_{NG}>1$ for low-concentration halos, consistent with the results of our N-body simulations (see figures~\ref{fig:EdS_bphi_vs_b0_all_nu},~\ref{fig:LCDM_bphi_vs_b0_all_nu}) as well as previous studies~\cite{Barreira:2022sey,Hadzhiyska:2025rez} of the $b_{h}-b_{\phi}$ relation exhibited by concentration-selected halos.

It is worth noting that the qualitative model presented in subsection~\ref{sssec:Low mass halos selected by concentration}, while consistent with our results from N-body simulations, is based on a number of simplifying assumptions which are nonetheless physically motivated and follow the spirit of heuristic models that have proven effective in capturing key aspects of halo formation and clustering~\cite{Asgari:2023mej,Cooray:2002dia}. A rigorous quantitative examination of these assumptions is required to fully validate the model and to assess its potential for making reliable, quantitative predictions. We plan to investigate along these lines in the future. 

Unlike low-mass ($M\ll M_{*}$) and high-mass ($M\gg M_{*}$) halos, the concentration of intermediate-mass ($M\sim M_{*}$) halos depends significantly on both the lagrangian properties of the progenitor peak (like the peak curvature) as well as the tidal shear caused by the gravitational influence of a neighbouring massive peak~\cite{Dalal:2008zd,Ramakrishnan:2019wtt} -- make the non-universality observed in intermediate-mass halos a combination of the two effects. A complete calculation of $p_{NG}$ as a function of concentration for \textit{all} halo masses using the model proposed in subsection~\ref{ssec:Halos selected by mass and concentration} is beyond the scope of this paper. However, in subsection~\ref{sssec:Intermediate mass halos selected by concentration}, we obtain a quantitative estimate for $p_{NG}$ as a function of halo peak-heights in a $\Lambda\text{CDM}$ cosmology (see figure~\ref{fig:pNG_Paranjape_Diemer}) assuming a log-normal concentration-mass PDF, together with the fitting formulas for the median concentration derived by~\citet{Diemer:2018vmz} and the Separate Universe responses of the mean and variance of the log-concentration derived by~\citet{ParanjapeSU}. While computed for the cosmology in \citet{ParanjapeSU}, different from the one used for our $\Lambda\text{CDM}$ simulations, the estimates of $p_{NG}$ obtained in subsection~\ref{sssec:Intermediate mass halos selected by concentration} do nevertheless correctly capture the trend in the $b_{h}-b_{\phi}$ relation of concentration-selected halos observed in our $\Lambda\text{CDM}$ simulations (see figure~\ref{fig:LCDM_bphi_vs_b0_all_nu}).

An important result of this paper is the observation that universality is restored at high halo masses in matter-dominated universes -- simply because such halos dominate their immediate environment and their properties are most tightly correlated with the lagrangian properties of their respective progenitor peaks (see subsection~\ref{sssec:High mass halos selected by concentration}). This result is observationally relevant because (a) high-mass halos observed in typical $\Lambda\text{CDM}$ cosmologies live in a matter-dominated universe and would, in principle show the same trend towards universality in the $b_{\phi}-b_{h}$ relation as for high-mass halos in a matter-dominated universe, and (b) the peak-height beyond which universality is restored for such halos is $\nu\sim 2.25$ which, in typical $\Lambda\text{CDM}$ cosmologies corresponds to a halo mass of $\sim 10^{12}-10^{13}\ M_{\odot}/h$ at redshifts $z\sim 3$. This falls within the range of quasar/LBG halo masses at redshift $z\sim 3$~\cite{Kim:2007de,Ishikawa_2017,Shen:2006ti}.

In realistic galaxy surveys, the observed galaxies are selected based on their observed physical properties and not on the properties of their respective host halos. These `selection function' effects can introduce a non-universality in the clustering of observed galaxies. In general, one can analyse selection function effects on the non-universality of observed galaxies using the same Separate Universe framework as presented in this paper -- by investigating the responses of the galaxy-halo connection to a background long-wavelength matter density perturbation and an increase in the amplitude of the primordial power spectrum.  For example, one could investigate the responses of halo-occupation distributions (HODs) to long-wavelength perturbations~\cite{Voivodic:2020bec} or the responses of stellar-to-halo-mass ratios (SHMRs)~\cite{Barreira:2020kvh}. To see how selection function effects can lead to non-universality and how they may be analysed using the Separate Universe framework, consider the simplest case of tracer abundance $n_{g}$ given in the halo-occupation-distribution (HOD) framework~\cite{HOD_1} by 
\begin{eqnarray}
    n_{g} = \int dM_{h}\ n_{h}(M)\langle N_{g}(M)\rangle\label{HOD}\ ,
\end{eqnarray}
where $n_{h}$ is the halo mass function (and is therefore universal in the sense of Eq.~\eqref{universality_statistics}) and $\langle N_{g}(M)\rangle$ is the global HOD number which encodes the mean number of tracers in a halo of mass $M$. The global HOD number $\langle N_{g}(M)\rangle$ could in principle depend on quantities which are \textit{non-universal} (in the sense that they don't obey Eq.~\eqref{universality_statistics}). Following the formalism of section~\ref{sec:Halo bias and the Separate Universe}, one can parametrise the non-universality in the abundance of such tracer (given by Eq.~\eqref{HOD}) in terms of $p_{NG}$ which yields :
\begin{eqnarray}
    p_{NG} &=& 1 + \frac{1}{n_{g}}\int dM_{h}n_{h}(M)\langle N_{g}(M)\rangle\times\nonumber\\
    &&\left(\frac{\partial\log \langle N_{g}\rangle}{\partial\delta_{L}}-2\delta^{-1}_{c}\cdot\frac{\partial\log\langle N_{g}\rangle}{\partial\log A_{s}}\right)\ .
\end{eqnarray}
The above equation shows that the non-universal nature of $\langle N_{g}(M)\rangle$ leads to $p_{NG}$ being different from unity.\footnote{Indeed,~\citet{Voivodic:2020bec} have computed the responses of HOD parameters to $A_{s}$ and $\delta_{L}$ and have shown that these responses are an important ingredient in modelling galaxy biases.} Similarly, ~\citet{Barreira:2020kvh} find that the non-universality of stellar-mass selected galaxies can be explained by the differing response of the median SHMR to a background, long-wavelength matter density and to an increase in the amplitude of the primordial power spectrum (caused by LPnG). In this context, our results in Section~\ref{ssec:Results for Halos selected by mass and concentration} suggest that any non-universality observed in high-bias, high-redshift galaxies (as against dark-matter halos) is an artifact of their underlying selection function and not of any intrinsic non-universality in the clustering of their host halos. We leave a detailed investigation of selection-function effects to future work.

\section*{Acknowledgements}

We are immensely grateful to Prof.\ Matthew McQuinn for insightful discussions and early collaborations. We would also like to thank Prof.\ Neal Dalal for helpful conversations and Benedikt Diemer for useful correspondence.
CS thanks Prof.\ Aseem Paranjape and Prof.\ Ashley Ross for helpful discussions.
CS and CH acknowledge support from the David and Lucile Packard Foundation; NASA grant 22-ROMAN11-0011; and the SPHEREx project under a contract from the NASA/Goddard Space Flight Center to the California Institute of Technology. CS also acknowledges support from NASA ATP award 80NSSC20K0541.
ML is supported by the NASA ATP award 80NSSC24K0937, Department of Energy grant DE-SC0011637 and the Dr. Ann Nelson Endowed Professorship of Physics. CS and DJ thank Eiichiro Komatsu for helpful comments on a draft.

\appendix
\section{Responses}

In this appendix we outline how the responses of the variance of the smoothed peak curvature $\sigma^{2}_{\xi,R} = \langle(R^{2}\nabla^{2}\delta_{R}^{2})^{2}\rangle$ and the cross-correlation coefficient $\gamma_{R} = \langle\delta_{R}\cdot R^{2}\nabla^{2}\delta_{R}\rangle/\sigma_{R}\sigma_{\xi,R}$ can be computed and also show that they obey Eq.s~\eqref{gamma_r_responses} and~\eqref{peak_curvature_responses}. We make use of standard results about the Separate Universe construction (as described in~\cite{Hu:2016ssz,Shiveshwarkar:2020jxr,Chiang:2018laa}). 

At the outset, we note that the variances $\sigma^{2}_{R}$, $\sigma^{2}_{\xi,R}$ and the cross-correlation $\Gamma_{R}=\langle\delta_{R}\cdot R^{2}\nabla^{2}\delta_{R}\rangle$ are all smoothed on a comoving scale R corresponding to the lagrangian radius encompassed by a halo mass $M$. In what follows, we suppress redshift-dependence wherever convenient and use $R$ and $M$ interchangeably with the understanding that $R$ is the lagrangian radius of the mass scale $M$ defined in the \textit{global} universe. The quantities $\sigma^2_{R}$, $\sigma^{2}_{\xi, R}$ and $\Gamma_{R}$ can be written as\footnote{We suppress the redshift dependence for convenience.} :
\begin{eqnarray}
    \sigma^{2}_{R} &=& \int \frac{d^{3}\textbf{k}}{(2\pi)^{3}}|\tilde{W}(kR)|^{2}\cdot P_{mm}(k)\ ,\label{sigma2R}\\
    \sigma^{2}_{\xi,R} &=&\int \frac{d^{3}\textbf{k}}{(2\pi)^{3}}(kR)^{4}|\tilde{W}(kR)|^{2}\cdot P_{mm}(k)\ ,\label{sigma2xiR}\\
    \Gamma_{R} &=& -\int\frac{d^{3}\textbf{k}}{(2\pi)^{3}}(kR)^{2}|\tilde{W}(kR)|^{2}\cdot P_{mm}(k)\ ,\label{GammaR}
\end{eqnarray}
where $P_{m}(k)$ is the linear CDM+Baryon matter power spectrum (at comoving wavenumber $k$, and $\tilde{W}(kR)$ is the Fourier transform of the smoothing window function which is typically chosen to a normalised top-hat profile in position space. From Eq.s~\eqref{sigma2R},~\eqref{sigma2xiR}, and~\eqref{GammaR}, we see that the variances $\sigma^2_{R},\ \sigma_{\xi,R}^2$ and the cross-correlation $\Gamma_{R}$ are all proportional to the amplitude of the primordial power spectrum $A_{s}$ implying 
\begin{eqnarray}
    2\frac{\partial\log\sigma_{R}}{\partial\log A_{s}} = 2\frac{\partial\log\sigma_{\xi,R}}{\partial\log A_{s}} = \frac{1}{\Gamma_{R}}\frac{\partial\Gamma_{R}}{\partial\log A_{s}}=1\ .\label{response_to_As}
\end{eqnarray}
For the cross-correlation coefficient $\gamma_{R} = \Gamma_{R}/\sigma_{R}\sigma_{\xi,R}$, Eq.~\eqref{response_to_As} implies 
\begin{eqnarray}
    \frac{\partial\gamma_{R}}{\partial\log A_{s}} = 0\ ,
\end{eqnarray}
as claimed in Eq.~\eqref{gamma_r_responses}.
For the Separate Universe responses, we note that the responses of $\sigma_{R}$, $\sigma_{\xi,R}$ and $\Gamma_{R}$ are completely determined by the response of the linear matter power spectrum $P_{mm}(k)$ to a background long-wavelength matter density fluctuation. The Separate Universe response of the matter power spectrum (in comoving coordinates) is~\cite{Shiveshwarkar:2020jxr,Chiang:2018laa}
\begin{eqnarray}
    \frac{\Delta P_{mm}(k)}{P_{mm}(k)\Delta\delta_{L}} = R_{growth}(k,z) + R_{geom}(k,z)\ ,\label{Pk_response} 
\end{eqnarray}
where $R_{growth}(k,z)$ encodes the effect of change in the matter power spectrum due to differing growth rates in the Separate Universe and the global universe and $R_{geom}(k,z)$ is a geometric factor that arises due to the difference in comoving coordinates\footnote{Note that the separate and global universes by construction have the same \textit{proper} coordinates -- which in general correspond to different comoving coordinates in the separate and global universes. For more details on the Separate Universe construction, see~\cite{Hu:2016ssz}.} as well as the different mean densities w.r.t. which density contrasts are defined in the Separate Universe and the global universe~\footnote{Note that density fluctuations in a Separate Universe are defined w.r.t a mean matter density of $\overline{\rho}_{m,SU} = \overline{\rho}_{m}(1+\delta_{L})$ where $\overline{\rho}$ is the mean matter density in the global universe and $\delta_{L}$ is the background long-wavelength mode within which the Separate Universe is constructed~\cite{Hu:2016ssz}.}

For a comoving scale $R$ corresponding to a `physical' scale (i.e., a fixed scale in proper coordinates, or a fixed halo mass $M$), it can be shown that the geometric contribution to the power spectrum response (i.e. $R_{geom}$ in Eq.~\eqref{Pk_response} \textit{does not} contribute to the Separate Universe responses of the variances $\sigma^{2}_{R}$, $\sigma_{\xi,R}^{2}$ and the cross-correlation $\Gamma_{R}$. This because the same `physical scale' corresponds to different comoving scales in the Separate and global universes\footnote{Note that the lagrangian radius of a halo of mass $M$ in the Separate Universe is different from the lagrangian radius in the global universe by a factor of $1-\frac{1}{3}\delta_{L}$ due to the different mean densities in the separate and global universes.}. Moreover, the window function $W$ which integrates to unity when integrated over global comoving coordinates does not integrate to unity over the comoving coordinates of the Separate Universe due to the difference in the comoving volume elements in both the separate and the global universes. Taking both these effects into account eliminates the effect of $R_{geom}$ (in Eq.~\eqref{Pk_response}) on the Separate Universe responses of $\sigma_{R}$, $\sigma_{\xi,R}$ and $\Gamma_{R}$\footnote{For more details on how this cancellation works, refer to~\cite{dePutter:2015vga}.}. That is why the Separate Universe response of $\sigma_{R}$, $\sigma_{\xi, R}$ and $\Gamma_{R}$ only arises from the growth term (i.e. $R_{growth}$) in Eq.~\eqref{Pk_response}. 

In matter-dominated cosmologies, the growth term $R_{growth}$ in Eq.~\eqref{Pk_response} can be shown to be constant at all redshifts and scales~\cite{Chiang:2014oga, Hu:2016ssz} and is equal to $R_{growth, \text{EdS}} = 26/21$. For typical $\Lambda\text{CDM}$ cosmologies, $R_{growth}$ is the same as in a matter-dominated cosmology ($=26/21$) till the end of the matter-dominated era and is weakly redshift-dependent (and still scale-independent to leading order) thereafter~\cite{Hu:2016ssz}. Plugging these values into Eq.~\eqref{sigma2R}, one can show that the Separate Universe response of $\sigma_{R}$ (to leading order) is
\begin{eqnarray}
    R_{SU} =\frac{\partial\log\sigma_{R}}{\partial\delta_{L}} = \frac{1}{2}\frac{1}{\sigma^{2}_{R}}\frac{\partial\sigma^{2}_{R}}{\partial\delta_{L}} = \frac{1}{2}R_{growth}\ ,
    \label{sigmaR_response}
\end{eqnarray}
where $R_{growth}\sim 26/21$ and is scale-independent and (for typical $\Lambda\text{CDM}$ cosmologies) weakly redshift-dependent at late times. This implies that $R^{-1}_{SU}\sim 1.6\sim \delta_{c}$ as mentioned in subsection~\ref{ssec:Halos selected by mass}.

In a similar way, one can show (after accounting for the difference in comoving coordinates in the separate and global universes) that to leading order,
\begin{eqnarray}
    \frac{\partial\log\sigma_{\xi,R}}{\partial\delta_{L}} = \frac{1}{2}\frac{1}{\sigma_{\xi,R}^{2}}\frac{\partial\sigma_{\xi,R}^{2}}{\partial\delta_{L}} = \frac{1}{2}R_{growth} = R_{SU}\ ,\label{sigmaxiR_response}
\end{eqnarray}
and 
\begin{eqnarray}
    \frac{1}{\Gamma_{R}}\frac{\partial\Gamma_{R}}{\partial\delta_{L}} = R_{growth} = 2\times R_{SU}\ ,\label{GammaR_response}
\end{eqnarray}
where $R_{SU}$ is the Separate Universe response of $\sigma_{R}$ (see equation~\eqref{sigmaR_response}). Eq.s~\eqref{sigmaR_response},~\eqref{sigmaxiR_response},~\eqref{GammaR_response} imply for the cross-correlation coefficient $\gamma_{R} = \Gamma_R/\sigma_{R}\sigma_{\xi,R}$ that, to leading order, 
\begin{eqnarray}
    \frac{\partial\gamma_{R}}{\partial\delta_{L}} = 0\ ,
\end{eqnarray}
as claimed in Eq.~\eqref{gamma_r_responses}. 
Moreover, Eq.s~\eqref{sigmaxiR_response} and Eq.~\eqref{response_to_As} imply that $\sigma_{\xi,R}$ obeys 
\begin{eqnarray}
    \frac{\partial\log\sigma_{\xi,R}}{\partial\delta_{L}} = 2\frac{\partial\log\sigma_{\xi,R}}{\partial\log A_{s}}R_{SU}(M,z)\ ,
\end{eqnarray}
as claimed in Eq.~\eqref{peak_curvature_responses}.
\label{app:responses}

\section{Binding energy associated with Hubble flow}
In this appendix, we compute the contribution of the background expansion of the unperturbed FRW universe to the binding energy of a low-mass peak which lies in proximity to a high-mass peak, when computed in Fermi coordinates. As mentioned in sec.~\ref{sssec:Low mass halos selected by concentration}, we construct Fermi coordinates centered at the trajectory of the high-mass peak and treat the motion of the low-mass peak (around its high-mass neighbour) in the non-relativistic, Newtonian limit. 

To begin with, consider the FRW metric in Fermi coordinates ($\textbf{x}_{F},t_{F})$~\cite{FNC2,FNC3} (to quadratic order in $Hx_{F}$) :

\begin{eqnarray}
    \label{eq:FRW_metric_in_FNC}
    ds^{2} &=& -\bigg[1-\left(\dot{H}+H^{2}\right)\textbf{x}_{F}^{2}\bigg]dt_{F}^{2}\nonumber\\
    &&+ \bigg[1-\frac{1}{2}H^{2}\textbf{x}^{2}_{F}\bigg]d\textbf{x}^{2}_{F}\ , 
\end{eqnarray}
where $H$ is the Hubble rate and we implicitly assume that the high-mass peak is at the origin $\textbf{x}_{F}=\textbf{0}$ of the Fermi coordinates. Note that by construction, the high-mass peak continues to remain at the spatial origin $\textbf{x}_{F}=\textbf{0}$ of the Fermi coordinates~\cite{FNC2,FNC3} for all times $t_{F}$.  

To quadratic order in $Hx_{F}$, the Fermi coordinates $(\textbf{x}_{F},t_{F})$ are related to the comoving coordinates $(\textbf{x}_{c},t_{c})$ as follows~\cite{FNC2,FNC3} :
\begin{eqnarray}
    \textbf{x}_{c} &=& \frac{\textbf{x}_{F}}{a(t)}\bigg[1 + \frac{1}{4}H^{2}\textbf{x}_{F}^{2}\bigg] \label{Fermi_to_comoving_x}\ ,\\
    t_{c} &=& t_{F} - \frac{1}{2}H\textbf{x}_{F}^{2}\label{Fermi_to_comoving_t}\ .
\end{eqnarray} 
The form of the metric in Eq.~\eqref{eq:FRW_metric_in_FNC} indicates that for non-relativistic motion at small distances (compared to the Hubble radius) the Hubble expansion in Fermi coordinates is encoded in a gravitational potential~\cite{Baumann:2010tm}
\begin{eqnarray}
    \label{phi_FRW}
    \Phi_{\rm FRW} = -\frac{1}{2}\left(\dot{H}+H^{2}\right)\textbf{x}_{F}^{2}\ .
\end{eqnarray}
For a matter-dominated universe, we have $\dot{H} = -\frac{3}{2}H^{2}$ leading to 
\begin{eqnarray}
    \label{phi_EdS_FRW}
    \Phi_{\rm FRW}\bigg|_{\rm EdS} = \frac{1}{4}H^{2}\textbf{x}_{F}^{2}\ .
\end{eqnarray}
Going further, Eq.s~\eqref{Fermi_to_comoving_t} and~\eqref{Fermi_to_comoving_t} imply that the Hubble flow (characterised by \textit{fixed} $\textbf{x}_{c}$) corresponds to a radial velocity field in the Fermi coordinates given (to quadratic order in $Hx_{F}$) by 
\begin{eqnarray}
    \label{vFRW}
    \textbf{v}_{\rm FRW} = \frac{d\textbf{x}_{F}}{dt_{F}}\bigg|_{\textbf{x}_{c}} = H\textbf{x}_{F}\ .
\end{eqnarray}
Eq.s~\eqref{phi_EdS_FRW} and~\eqref{vFRW} show that in the non-relativistic limit, the Hubble flow in Fermi coordinates corresponds to a binding energy $\mathcal{E}_{\rm FRW}$ given by 
\begin{eqnarray}
    \mathcal{E}_{\rm FRW} = \Phi_{\rm FRW} + \frac{1}{2}\textbf{v}_{\rm FRW}^{2}\ . 
\end{eqnarray}
In a matter-dominated universe, this becomes (following Eq.~\eqref{phi_EdS_FRW}) 
\begin{eqnarray}
    \mathcal{E}_{\rm FRW} = \frac{1}{4}H^{2}\textbf{x}_{F}^{2} + \frac{1}{2}H^{2}\textbf{x}_{F}^{2} = \frac{3}{4}H^{2}\textbf{x}_{F}^{2}\ ,
\end{eqnarray}
which is the binding energy associated with the Hubble flow in Fermi coordinates and is the leading order term in Eq.~\eqref{FNC_binding_energy}. 

In the presence of a perturbed FRW space-time, the metric in Fermi coordinates centered at the high-mass peak will also acquire corrections on top of the zeroth order form associated with the Hubble flow (given by Eq.~\eqref{eq:FRW_metric_in_FNC}). These corrections are proportional to the space-time perturbations in the FRW metric and their derivatives~\cite{LSS_Shear,FNC2,FNC_1}. In particular, both the gravitational potential $\Phi_{\rm FRW}$ and the velocity field $\textbf{v}_{\rm FRW}$ associated with the Hubble flow (given by Eq.s~\eqref{phi_FRW} and~\eqref{vFRW}) acquire perturbations 
\begin{eqnarray}
    \label{phi_perturbed_FRW}
    \Phi_{\rm net} &=& \Phi_{\rm FRW} + \Delta\Phi\nonumber\ ,\\
    \label{v_perturbed_FRW}
    \textbf{v}_{\rm net} &=& \textbf{v}_{\rm FRW} + \Delta\textbf{v}\ ,
\end{eqnarray}
with the binding energy acquiring perturbations coming from $\Delta\Phi$ and $\Delta\textbf{v}$
\begin{eqnarray}
    \mathcal{E} = \mathcal{E}_{\rm FRW} + \Delta\mathcal{E}\ ,
\end{eqnarray}
where $\Delta\mathcal{E} \sim \Delta\Phi + \textbf{v}_{\rm FRW}\cdot\Delta\textbf{v}$.

While deriving explicit expressions for the fluctuations $\Delta\Phi$ and $\Delta\textbf{v}$ is beyond the scope of this paper, what we can say is that they are suppressed in magnitude by a factor of $A^{1/2}_{s}D$ (where $D$ is the linear growth factor and is no more than $\mathcal{O}(1)$ in magnitude) compared to the FRW terms $\Phi_{\rm FRW}$ and $\textbf{v}_{\rm FRW}$ -- implying that the response $\partial\mathcal{E}/\partial\log A_{s}$ is $\sim\mathcal{O}(A^{1/2}_{s}D)$.

\label{app:FNC}
\newpage
\bibliographystyle{apsrev4-1}
\bibliography{references.bib}

\end{document}